\documentclass[12pt]{article}
\usepackage{amsmath,amssymb}
 \usepackage{graphicx}

\usepackage{graphicx}
\usepackage{wrapfig}
\usepackage{color}

\usepackage{hyperref}
\usepackage[sort,compress]{cite}

\def\beq{\begin{eqnarray}}
\def\eeq{\end{eqnarray}}

\def\lb{\label}

\newcommand{\be}{\begin{equation}}
\newcommand{\ee}{\end{equation}}
\newcommand{\bea}{\begin{eqnarray}}
\newcommand{\eea}{\end{eqnarray}}
\newcommand{\bg}{\begin{gather}}

\newcommand{\bseq}{\begin{subequations}}
\newcommand{\eseq}{\end{subequations}}

\renewcommand{\ln}{\mathop{\rm ln}\nolimits}

\def\be{\begin{eqnarray}}
\def\ee{\end{eqnarray}}
\def\lb{\label}

\begin{document}

\title{\textbf{
Quantum states and their back-reacted geometries in 2d dilaton gravity}}
\vspace{0.5cm}
\author{ \textbf{ Yohan Potaux$^{1}$, Debajyoti Sarkar$^{2}$ and Sergey N. Solodukhin$^{1}$ }} 

\date{}
\maketitle
\begin{center}
    \emph{$^{1}$Institut Denis Poisson UMR 7013,
  Universit\'e de Tours,}\\
  \emph{Parc de Grandmont, 37200 Tours, France} \\
\vspace{0.4cm}
\emph{$^{2}$Department of Physics}\\
  \emph{Indian Institute of Technology Indore}\\
  \emph{Khandwa Road 453552 Indore, India}
\end{center}




\vspace{0.2mm}

\begin{abstract}
\noindent{Within the Russo-Susskind-Thorlacius (RST) two-dimensional model that includes a scalar (dilaton) field we address the important question  of how the classical black hole geometry is modified in a semiclassical gravitational theory. It is the principle goal of this paper to analyze what is the back-reacted geometry that corresponds to a given quantum state. The story is shown to be dramatically different for the Hartle-Hawking state (HH) and for the Boulware state. In the HH case the back-reacted geometry is a modification of the classical black hole metric that still has a smooth horizon with a regular curvature. On the other hand, for the Boulware state the classical horizon is replaced by a throat in which the $(tt)$ component of the metric (while non-zero) is extremely small. The value of the metric at the throat is bounded by the inverse of the classical black hole entropy. On the other side of the throat the spacetime is ended at a null singularity. More generally, we identify a family of quantum states and their respective back-reacted geometries. We also identify a certain duality in the space of states. Finally, we study a hybrid set-up where both physical and non-physical fields, such as the ghosts, could be present. We suggest that it is natural to associate ghosts with the Boulware state, while the physical fields can be in any quantum state. In particular, if the physical fields are in the HH state, then the corresponding semiclassical geometry is horizonless. Depending on the balance between the number of physical fields and ghosts, it generically has a throat that may join with another asymptotically flat region on the other side of the throat.
}
\end{abstract}


\rule{7.7 cm}{.5 pt}\\
\noindent ~~~ {\footnotesize   yohan.potaux@lmpt.univ-tours.fr, dsarkar@iiti.ac.in, Sergey.Solodukhin@lmpt.univ-tours.fr}

\pagebreak

\section{Introduction}
\setcounter{equation}0
\bigskip
Black holes are interesting geometric objects that typically appear as solutions to the classical gravitational equations.
The existence of the horizons is the main  property that defines the black hole spacetime in any dimension.
In the simplest non-rotating case the black hole horizon is defined by the condition that the $(tt)$ component of the metric vanishes.
As is well known, the horizon leads to some very peculiar properties. Classically, the part of the spacetime  
that is inside the horizon becomes inaccessible for any outside observer. 

A step towards non-classicality will be to add quantum fields, considering them on the black hole background that still solves the classical gravitational equations.
In this picture the horizon does not appear to be  absolutely opaque. Instead, there appears radiation from the black hole that is seen as a thermal radiation 
at the Hawking temperature by an asymptotic observer. In fact, the presence of this thermal radiation is known to depend on the choice of the states of the quantum fields. 
Some of the quantum states that have been discussed in the literature are as follows.

\medskip

\noindent{\it The Hartle-Hawking state}:  it contains  thermal radiation at infinity and the stress-energy  tensor is regular at the horizon. It describes a black hole in thermal equilibrium with the Hawking radiation.
 
  \vspace{5mm}
  
\noindent {\it The Boulware state}: the stress-energy tensor is  vanishing  at infinity and there is no radiation there. However, it is singular at the horizon.
 
  \vspace{5mm}
 
\noindent {\it The Unruh state}:  the stress-energy tensor is regular only at the future horizon, and there is a thermal flux
of radiation at future null infinity. It describes the process of the black hole evaporation.

Generally, the situation appears to be similar to the choice of boundary conditions in an open domain: fixing the regularity condition at one end, one gets a singular behaviour at the other.
In this paper we prefer to start from the asymptotic infinity, where we impose the conditions. What happens at the horizon is then just a consequence of this choice.
Practically, the Unruh state is perhaps the most physically justified quantum state. However, it corresponds to a time evolving situation and will not be considered here any further, since we focus on the static case only.

A natural question  that was addressed in the literature is  how the respective quantum stress-energy tensors back-react on the geometry.
There have been quite a few papers on this subject, see  for instance \cite{York:1986it}-\cite{Zaslavskii:2003is}  that used certain approximations  for the quantum stress tensor.\footnote{The other interesting approach is related to brane paradigm \cite{Emparan:2002px} that will not be discussed here.}
However in those papers, at most one was able to compute the
stress tensor on a given solution to the classical gravitational equations. Whereas, what one wants to know is its form for a generic black hole metric.

Since the classical gravity is necessarily modified in the quantum theory, it is clear that the black holes should somehow be embedded
in a more general quantum gravitational theory. The latter is yet to be properly defined. A good approximation to such a theory is 
given by the so-called semiclassical gravitational theory. Indeed, each quantum field propagating on a spacetime background will
produce a modification to the classical gravitational theory. In this approach the metric is still classical, even though the modifications due to quantum gravitons
(small quantum perturbations over the classical background)  can also be considered. The semiclassical gravitational theory is a rather complicated theory that contains both local and non-local terms. 
Some of such non-local terms were computed in a series of papers  \cite{Barvinsky}  where the gravitational effective action was expanded in powers of the curvature, with the coefficients in the expansion
being the non-local form-factors. In four dimensions and in the cubic order one counts as many as 29 invariants as was shown in \cite{Barvinsky}.
Thus, it is clearly a rather complicated problem to be addressed in four dimensions.

Taking the complexity of the problem what are the basic questions we would like to answer? Here are some of such questions:


\begin{itemize}

\item does the quantum-corrected metric have a horizon?


\item  if there is a horizon, how does its position and the Hawking temperature change with respect to the classical situation?


\item what happens at asymptotic infinity?  If there exists a thermal Hawking radiation as in the Hartle-Hawking state,  then it curves the spacetime and it is likely that we no longer have Minkowski spacetime.


\item how do the answers to the previous questions depend on the choice of the quantum state?  What are the back-reacted geometries for the Hartle-Hawking and Boulware states?

\item provided the quantum-corrected geometries are horizonless, how close are they to the classical black hole geometry? Can they be considered as the black hole mimickers?

\end{itemize}


We stress that our main objective is the four-dimensional case where some particular understanding has recently been achieved \cite{Berthiere:2017tms}. Due to the complexity of  the
gravitational effective action in four dimensions, it is a good idea to analyse the problem in a somewhat simplified while still meaningful setting.
In the present paper we are going to address all these questions in a semiclassical model of two-dimensional dilaton gravity. In two dimensions the gravitational effective action is given by the Polyakov action provided the quantum matter is a conformal field theory. The simplicity of the theory makes it a quite attractive toy model  that motivated a very active study in the 90's. This direction of research was initiated in 
\cite{Callan:1992rs},  for a review  see \cite{Strominger:1994tn}.
More specifically we shall study the model proposed by Russo, Susskind and Thorlacius (RST) \cite{RST}. The important advantage of this model is that it is exactly integrable similar to the classical dilaton gravity.  For the Hartle-Hawking state it was fully analysed in \cite{Solodukhin:1995te}.
As we show in the present paper this integrability can be extended further to any quantum state including the Boulware state.\footnote{When the present study was in the final stages, we have learned about an earlier paper by Zaslavsky \cite{Zaslavskii:2006pn} which addressed some of the aspects relevant to our discussion of the Boulware state in section \ref{sec:boul}. }

A short outline of our paper is as follows. After briefly reviewing the classical dilaton action in section \ref{sec:classdilgrav}, in section \ref{sec:classandpolyakov} we discuss the various states that we obtain once we add the Polyakov action corresponding to conformal quantum matter. Some of the main outcomes of this section is to point out a one-parameter family of states besides Hartle-Hawking (HH) and Boulware, and to also uncover an interesting duality connnecting two HH states in the space of these parameter values. In section \ref{sec:rst}, we briefly review the two types of solutions that one obtains from the field equations corresponding to the RST model. In particular, in this paper we focus on the non-constant dilaton solution. In section \ref{sec:hh}, we study the resulting quantum geometry in HH state and provide the asymptotic spacetime solution due to the presence of the Hawking radiation. In section \ref{sec:solvegeneq}, we solve the complete effective action for arbitrary states and obtain the master equations that we use in the following parts of the paper. Section \ref{sec:boul} is devoted towards the study of Boulware vacuum, and its corresponding subsections explore various different limits and cases leading to geometries with wormhole-type throat structures, null singularities etc. In section \ref{sec:genC} we deal with the completely general one-parameter family of states mentioned earlier. Section \ref{sec:hybrid} studies the hybrid case of physical fields and ghosts and their backreaction to the geometry. Finally we conclude in section \ref{sec:conclusions} discussing several implications of our results and potential future directions.

\section{Black holes in classical dilaton gravity}\label{sec:classdilgrav}
\setcounter{equation}0
\bigskip

As is well known, the Einstein-Hilbert action in two dimensions does not produce any non-trivial equations.
So in order to introduce the gravitational dynamics in two dimensions one has to modify the gravitational action
either by considering certain non-linear functions of curvature $f(R)$, or stay in the class of theories with only two derivatives in the field
equations and introduce some additional fields. A rather standard way is to introduce a scalar field $\phi$, called dilaton. In the class of two-dimensional theories of
dilaton gravity the most popular is the so-called string inspired dilaton theory described by the action \cite{2d}
\be
I_0=\frac{1}{2\pi}\int_M d^2 x \sqrt{-g}e^{-2\phi}(R+4(\nabla\phi)^2+4\lambda^2)\, ,
\lb{1}
\ee
where we've omitted the possible boundary terms.

Variation of the action with respect to the metric leads to the gravitational equations
\be
T^{(0)}_{\mu\nu}\equiv \frac{1}{\pi}e^{-2\phi}(2\nabla_\mu\nabla_\nu\phi-2g_{\mu\nu}(\Box \phi-(\nabla\phi)^2+\lambda^2)=0\, .
\lb{2}
\ee
On the other hand the variation with respect to the field $\phi$ gives us the dilaton equation
\be
e^{-2\phi}(R+4\Box\phi-4(\nabla\phi)^2+4\lambda^2)=0\, .
\lb{3}
\ee
Looking separately at the trace and the trace-free  parts of (\ref{2}) and taking into account (\ref{3}), one arrives
at the following set of equations,
 \be
 \nabla_\mu\nabla_\nu \phi=\frac{1}{2}g_{\mu\nu}\Box\phi\, ,
 \lb{4}
 \ee
 \be
 R=-2\Box\phi\, ,
 \lb{5}
 \ee
 \be
 R+4(\nabla\phi)^2-4\lambda^2=0\, .
 \lb{6}
 \ee
A consequence of (\ref{4}) is that the vector field $\xi_\mu=\epsilon_\mu^{\ \nu}\partial_\nu \phi$ is the Killing vector, i.e.
$\nabla_\mu\xi_\nu+\nabla_\nu\xi_\mu=0$.  Its norm is equal to $\xi^2=-(\nabla\phi)^2$ so that vector $\xi$ is null at the critical points of $\phi$, i.e. $(\nabla\phi)^2=0$.
Along the Killing trajectories given by $\xi$ the dilaton $\phi$  is constant as $\xi^\mu\nabla_\mu\phi=0$. So that it is natural to choose $\phi$ as a space-like coordinate
and associate $\xi$ with a time coordinate, $\xi=\partial_t$.  The general solution to the field equations is thus a static metric of the form
\be
&&ds^2=-g(x)dt^2+g^{-1}(x)dx^2\, ,\nonumber \\
&&\phi=-\lambda x\,,  \ \  g(\phi)=1-ae^{2\phi}\, .
\lb{7}
\ee
For $a=0$ the metric is flat. This is the so-called linear dilaton vacuum.  For positive $a$ the metric describes an asymptotically flat space-time at $\phi\rightarrow -\infty$, and has a curvature singularity $R=-4\lambda^2 ae^{2\phi}$ at $\phi=+\infty$. The point $\phi=\phi_h$ where  the metric function vanishes, i.e. $g(\phi_h)=0$, is the Killing horizon.
Vector $\xi$ becomes null here, $\xi^2(\phi_h)=0$. For negative $a$ the horizon is absent and the solution describes a naked singularity.

The metric (\ref{7}) then describes a two-dimensional black hole with mass $M=\frac{\lambda a}{\pi}$. The Hawking temperature is independent of mass,
$T_H=\frac{\lambda}{2\pi}$. This appears to be a peculiarity of two dimensions. The entropy of the black hole is determined by the value of the dilaton field at the horizon,
$S_{BH}=2e^{-2\phi_h}=2a$. For a discussion of the thermodynamics of a classical 2d dilaton black hole see \cite{2d-2}.

Comparing this two-dimensional picture to the four-dimensional case we see that the dilaton, or more precisely $e^{-\phi}$, could be identified with the radial coordinate $r$. So that the entropy would have the usual interpretation as the ``area''.

\section{Vacua of quantum CFT on a 2d black hole background}\label{sec:classandpolyakov}
\setcounter{equation}0
\bigskip

Now we take a  step towards the quantum gravitational theory and consider quantum matter on the classical black hole background.
To make things simpler we consider a conformal field theory. The corresponding quantum effective action, provided the quantum fields are integrated out, 
is known to be the Polyakov action. The Polyakov action is a non-local functional of the background metric. We however prefer to deal with a local version of the action.
This always can be done by introducing an auxiliary field $\psi$,
\be
I_1=-\frac{\kappa}{2\pi}\int_M d^2 x\sqrt{-g}\,\left(\frac{1}{2}(\nabla\psi)^2+\psi R\right)\, ,
\lb{8}
\ee
where we again omit the possible boundary terms. For a multiplet of $N$ scalars one has $\kappa=\frac{N}{24}$. If one includes the ghosts then
$\kappa=\frac{N-24}{24}$.  The negative  number $-24$ comes out as $-24=-26+2$ when one quantizes the dilaton gravity, $-26$ being the contribution of the ghosts,
see \cite{Russo:1992yg}-\cite{Bilal:1992kv}.
 In the next sections, where we consider the back-reaction problem,  the parameter $\kappa$  will control the quantum 
modifications to the classical geometry. So that sometimes we will want to take the limit of very small $\kappa$ in order to illustrate how the semiclassical geometry approaches the classical one.
Thus, $\kappa$ will be treated as a continuous parameter.
It mostly takes positive values but its negative values may also be of some interest, which we have studied below in the context of various quantum states.
Its role, when the back-reacted geometry is  considered in two dimensions, is similar to the Newton's constant $G$ in four dimensions. In this section, however, $\kappa$
simply measures  the number of degrees of freedom in the quantum conformal field theory in question.

Variation of this action with respect to $\psi$ gives us 
\be
\Box \psi=R\, .
\lb{9}
\ee
This equation can be formally solved for $\psi$: $\psi(x)=\int dy\, G(x,y) R(y)$, where we introduced a Green's function, $\Box_x G(x, y)=\delta(x-y)$.
 The substitution of $\psi$ back to the action (\ref{8}) will lead to the  usual non-local version of the Polyakov action. As always, the Green's function is defined up to a solution to the homogeneous equation and hence one should specify the appropriate boundary conditions to uniquely define it. In other words, there is a freedom in defining $\psi$.
 In the dilaton gravity defined in section \ref{sec:classdilgrav} one has that $R=-2\Box\phi$ so that there exists a relation between $\psi$ and $\phi$,
 \be
 \psi=-2\phi+w\, ,
 \lb{10}
 \ee
 where $w$ solves the homogeneous equation
 \be
 \Box w=0\, .
 \lb{11}
 \ee
Variation of (\ref{8}) with respect to metric gives the stress-energy tensor for the quantum CFT,
\be
T^{(1)}_{\mu\nu}=-\frac{\kappa}{2\pi}\left(\partial_\mu\psi\partial_\nu \psi-2\nabla_\mu\nabla_\nu\psi -g_{\mu\nu}\left(-2R+\frac{1}{2}(\nabla \psi)^2\right)\right)\, .
\lb{12}
\ee
In a static two-dimensional metric of the form (\ref{7}), assuming that $\psi$ is only a function of coordinate $x$ (but not of time $t$), one finds
\be
T^{(1)0}_{\ \ \ 0}=\frac{\kappa}{2\pi}\left(\psi'(x)g'(x)+\frac{1}{2}g(x)\psi'^{2} (x)+2g''(x)\right)\, 
\lb{13}
\ee
for the energy density (note that for metric (\ref{7}), the scalar curvature is $R=-g''(x)$). On the other hand, the homogeneous equation (\ref{11}) can be solved as
(again we assume here  that $w$ does not depend on time $t$)
\be
w'(x)=\frac{C}{g(x)}\, ,   \   \   w(x)=C\int^x \frac{dy}{g(y)}\, .
\lb{14}
\ee
Here  $C$ is an integration constant. As we will soon see, this constant incorporates the information on the choice of the vacuum.
For the solution of the dilaton gravity considered in section \ref{sec:classdilgrav} one finds that (\ref{13}) takes a simple form
\be
T^{(1)0}_{\ \ \ 0}=\frac{\kappa}{\pi}\left(3\lambda^2g(x)-2\lambda^2+\frac{C(C+4\lambda)}{4g(x)}\right)\, .
\lb{15}
\ee
As a result, asymptotically, where $g(x)$ approaches $1$, the energy density turns out to be
\be
T^{(1)0}_{\ \ \ 0}=\frac{\kappa}{4\pi}(C+2\lambda)^2\, .
\lb{16}
\ee
Now we are ready to define the different quantum vacua.

\bigskip

\noindent\underline{\it The Hartle-Hawking (HH) state}

\medskip

By definition, the Hartle-Hawking state is the one that is regular at the horizon, i.e. $g(x_h)=0$. As follows from (\ref{15}) there are two values of $C$ for which the divergent term
in the energy density at the horizon vanishes:
\be
C=0\,  \  \  \  \   {\rm and}   \  \  \  C=-4\lambda \, .
\lb{17}
\ee
At infinity the energy density (\ref{16}) is then 
\be
T^{(1)0}_{\ \ \ 0}=\frac{\kappa\lambda^2}{\pi}=\frac{\pi}{6}N T^2_H\,,
\lb{18}
\ee
which is precisely the energy density of the radiation at the Hawking temperature $T_H=\frac{\lambda}{2\pi}$.
This is of course the expected behavior of the Hartle-Hawking state. It is known to describe the black hole in equilibrium with the thermal Hawking radiation.
For the range of the coordinate $x$ from the horizon $x_h$ to infinity the quantum energy density in the Hartle-Hawking state for the solution (\ref{7}) reads
\be
T^{(1)0}_{\ \ \ 0}=\frac{\kappa\lambda^2}{\pi} \left(1-3ae^{2\phi}\right)\,.
\lb{19-1}
\ee
At the horizon it takes some negative value $T^{(1)0}_{\ \ \ 0}=-\frac{2\kappa\lambda^2}{\pi}$.

\bigskip

\noindent\underline{\it The Boulware state}

\medskip

The Boulware state is defined by the condition that it is empty at asymptotic infinity, i.e. $T^{(1)0}_{\ \ \ 0}=0$
 when $g\rightarrow 1$. As is seen from (\ref{16}), this condition singles out the value 
 \be
 C=-2\lambda\, .
 \lb{19}
 \ee
 The energy density then is divergent at the horizon. This is the expected and in fact well-known property of the Boulware state.
In the region between the horizon and the asymptotic infinity the energy density 
\be
T^{(1)0}_{\ \ \ 0}=-\frac{\kappa\lambda^2}{\pi}a^{2\phi}\left(3+\frac{1}{1-ae^{2\phi}}\right)
\lb{20}
\ee
is everywhere negative.

\bigskip

\noindent\underline{\it A general $C$-state}

\medskip

In general there exists a family of quantum states, parametrized by $C$. For values of $C$ 
different from $0, -2\lambda$, $-4\lambda$ such a $C$-state would be an intermediate quantum state that shares certain properties of the 
Hartle-Hawking state and of the Boulware state. Such a $C$-state would not be empty at asymptotic infinity (as the Hartle-Hawking state)
and at the same time it would be divergent at the horizon (as the Boulware state).

To end this section, we want to emphasize that there exists an interesting duality in the space of quantum states parametrized by $C$.
The difference in the energy density for two values of $C$ is in the term which is divergent at the horizon,
\be
T^{(1)0}_{\ \ \ 0}(C_1)-T^{(1)0}_{\ \ \ 0}(C_2)=\frac{\kappa}{4\pi g(\phi)}(C_1-C_2)(C_1+C_2+4\lambda)\, .
\lb{21}
\ee
This indicates that the energy density is the same if $C_1$ and $C_2$ are related by relation
\be
C_1+C_2=-4\lambda\, .
\lb{22}
\ee
This explains why there are two values of $C$ for the Hartle-Hawking state. On the other hand, the Boulware state appears to be symmetric
under this duality. As far as we are aware, these arbitrary $C$-states and this duality in the $C$-parameter space were not noticed in any earlier literature. We also note that even though the energy density is the same for these two values of $C$, the function $w$ (and therefore $\psi$) is not.
The value of function $\psi$ at the horizon carries information about the entropy (see for instance \cite{Myers:1994sg,Solodukhin:1994yz}). As a result, there may still appear some important physical 
differences between these two HH states. However, we will not discuss this issue in the present paper.

\bigskip

\noindent\underline{\it Backreaction in asymptotic region}

\medskip

The non-vanishing stress-energy present at infinity, (\ref{16}) for a generic $C$, will necessarily curve the spacetime.
This will lead to some subleading terms in the metric present asymptotically, when $\phi\rightarrow -\infty$. 
In order to analyze the asymptotic geometry we take metric in the form
\be
ds^2=-g(\phi)dt^2+h^2(\phi)g^{-1}(\phi) dt^2\, 
\lb{b-1}
\ee
and assume that asymptotically one has that
\be
g=1+\delta g\, ,  \  \ h=-1/\lambda +\delta h\, ,
\lb{b-2}
\ee
where $\delta g$ and $\delta h$ are small  perturbations over the linear dilaton vacuum.
Equation $R+2\Box\phi=0$ will lead to a relation
\be
\delta h'=\frac{1}{2\lambda}(\delta g''-2\delta g')\, ,
\lb{b-3}
\ee
where all derivatives are with respect to dilaton $\phi$.
On the other hand, one finds
\be
\delta T^{(0)0}_{\ \ \ 0}=\frac{\lambda^2}{2\pi}e^{-2\phi}(-\delta g''+2\delta g')\, .
\lb{b-4}
\ee
The gravitational equations with a source in the form of (\ref{16}), $T^{(0)0}_{\ \ \ 0}+T^{(1)0}_{\ \ \ 0}=0$, then leads to equation\footnote{In the RST model, which is considered in the rest of the paper, we also have a third term $T^{(2)0}_{\ \ \ 0}$ due to the local term which is added to preserve a classical symmetry in the dilaton gravity action. Analysis shows that this term
$T^{(2)0}_{\ \ \ 0}=O(e^{2\phi})$ is subleading in the asymptotic region and hence it can be ignored in the asymptotic analysis.}
\be
\delta T^{(0)0}_{\ \ \ 0}+\frac{\kappa}{4\pi}(C+2\lambda)^2=0
\lb{b-5}
\ee
that solves as follows
\be
\delta g=\frac{\kappa}{4\lambda^2}(C+2\lambda)^2\phi e^{2\phi}\, .
\lb{b-6}
\ee
Using (\ref{b-3}) one finds
\be
\delta h=\frac{\kappa}{8\lambda^3}(C+2\lambda)^2 e^{2\phi}\, .
\lb{b-7}
\ee
Equations (\ref{b-6}) and (\ref{b-7}) present the modifications in the asymptotic geometry produced by the non-vanishing  stress energy tensor.
For $C=0$ (or $C=-4\lambda$) this corresponds to the backreaction of the thermal radiation on the spacetime metric.

\section{The RST model}\label{sec:rst}
\setcounter{equation}0
\bigskip

Now we would like to address the question of the back-reaction of a quantum state to the geometry. The appropriate two-dimensional model for this purpose is the
so-called RST model. It was suggested by Russo, Susskind and Thorlacius in 1992 \cite{RST}. Its important advantage is that it is exactly integrable. 
The integrability is related to the fact that one preserves  a certain symmetry present in the classical action.
The action of the model is a sum of three terms: the classical dilaton action $I_0$ (\ref{1}), the Polyakov action $I_1$ (\ref{8}) and a new local term $I_2$,
\be
I_{RST}=I_0+I_1+I_2\, ,  \  \  \ I_2=-\frac{\kappa}{2\pi}\int d^2 x \sqrt{-g}\phi R
\lb{23}
\ee
Varying this action with respect to metric one finds
\be
T_{\mu\nu}\equiv T^{(0)}_{\mu\nu}+T^{(1)}_{\mu\nu}+T^{(2)}_{\mu\nu}=0\, ,
\lb{24}
\ee
where we have previously defined $T^{(0)}_{\mu\nu}$ (\ref{2}) and $T^{(1)}_{\mu\nu}$ (\ref{12}). Variation of the last term $I_2$ in (\ref{23}) with respect to metric
gives
\be
T^{(2)}_{\mu\nu}=-\frac{\kappa}{\pi}\,\left(g_{\mu\nu}\,\Box\phi-\nabla_\mu\nabla_\nu\phi\right)\, .
\lb{25}
\ee
Variation of the total action with respect to dilaton $\phi$ gives the dilaton equation
\be
2e^{-2\phi}(R+4\Box\phi-4(\nabla\phi)^2+4\lambda^2)=-\kappa R\,.
\lb{26}
\ee
On the other hand, taking the trace of (\ref{24}) one gets
\be
2e^{-2\phi}(\Box\phi-2(\nabla\phi)^2+2\lambda^2)=-\kappa (R+\Box\phi)\, .
\lb{27}
\ee
Combining these two equations one arrives at a simple equation
\be
(R+2\Box\phi)(\kappa-2e^{-2\phi})=0\, .
\lb{28}
\ee
It has solutions of two different types. 

\bigskip

\noindent\underline{Anti-de Sitter space-time (constant dilaton)}

\medskip

The solution of first type is characterised by a constant value of the dilaton 
\be
\phi=-\frac{1}{2}\ln\frac{\kappa}{2}={\rm const}\, .
\lb{29}
\ee
It follows from either equation (\ref{26}) or (\ref{27}) that the scalar curvature is constant in this case,
\be
R=-2\lambda^2\, .
\lb{30-ads}
\ee
This is a two-dimensional anti-de Sitter space-time. Even though the value of the dilaton (\ref{29}) is ``quantum'' the value of the curvature
(\ref{30-ads}) is classical.   The existence of this solution was demonstrated in \cite{Solodukhin:1995te}.
This solution may be interesting by itself. We, however, will not consider it in the present paper.

\bigskip

\noindent\underline{Black hole type solution (non-constant dilaton)}

\medskip

The other solution is characterised by a varying dilaton. Then equation (\ref{28}) solves as follows
\be
R=-2\Box\phi\, .
\lb{30}
\ee
This is the same equation as in the classical case, see (\ref{5}). This essentially simplifies the integration of the equations.
Taking that the equation for the auxiliary field $\psi$ is still $\Box\psi=R$,  equation (\ref{30}) can be solved in the same way as before,
\be
\psi=-2\phi+w\, ,  \  \  \   \Box w=0\, .
\lb{31}
\ee
Further integration of this equation depends on the choice of function $w$. As we have discussed earlier in the paper, the choice of $w$ depends on the choice of the quantum state (or vacuum).

\section{Quantum-corrected black hole (the back-reacted Hartle-Hawking state)}\label{sec:hh}
\setcounter{equation}0
\bigskip

Our first choice is $w={\rm const}$ so that all derivatives of $w$ vanish. As we have seen  in section \ref{sec:classandpolyakov}, this choice corresponds to the Hartle-Hawking vacuum
of the  quantum conformal field theory. This case was analyzed in details in \cite{Solodukhin:1995te}. In this section we give a brief summary of findings made in
\cite{Solodukhin:1995te}.

With this choice of $w$, the trace-free part of (\ref{24}) can be presented in the form
\be
\nabla_\mu\nabla_\nu F(\phi)=\frac{1}{2}g_{\mu\nu}\Box F(\phi)\, ,
\lb{32}
\ee
where we introduced
\be
F(\phi)\equiv \phi-\frac{\kappa}{4}e^{2\phi}\, .
\lb{33}
\ee
This equation is similar to (\ref{4}) of the classical case. Similarly to what we had in the classical dilaton gravity, the equation (\ref{32})
implies that vector $\xi_\mu=\epsilon_\mu^{\ \nu}\partial_\nu F(\phi)$ is  the Killing vector. This fact essentially simplifies the integration of the equations.
The solution is the presented as follows
\be
&&ds^2=-g(\phi)dt^2+g^{-1}(\phi)h^2(\phi)d\phi^2\, , \nonumber \\
&&h(\phi)=-\frac{1}{\lambda}F'(\phi)=-\frac{1}{\lambda}\left(1-\frac{\kappa}{2}e^{2\phi}\right) \, , \nonumber \\
&&g(\phi)=1+\kappa\phi e^{2\phi}-a e^{2\phi} \, .
\lb{34}
\ee
This solution in the present form was found in \cite{Solodukhin:1995te}. It  represents a quantum modification (parametrized by $\kappa$) of the classical black hole metric (\ref{7}).
Asymptotically, when $\phi\rightarrow -\infty$, the metric function $g(\phi)$ approaches value $1$ with a correction term
$g(\phi)\simeq 1+\kappa \phi e^{2\phi}$, which is due to  the backreaction of the stress-energy of the thermal radiation
present in the Hartle-Hawking state (e.g. compare with \eqref{b-6} for $C=0$ or $-4\lambda$). The metric however remains asymptotically flat as the curvature vanishes as $\phi\rightarrow-\infty$.

The metric (\ref{34}) has a curvature singularity at $\phi=\phi_{cr}$  where $h(\phi_{cr})$ vanishes,
\be
R=-\frac{4 \lambda^{-1}e^{2\phi}}{h(\phi)^3}\,\left(a-\kappa-\kappa\phi+\frac{\kappa^2}{4}e^{2\phi}\right)\, .
\lb{34-1}
\ee
 One finds that $\phi_{cr}=-\frac{1}{2}\ln\frac{\kappa}{2}$.
The value of the metric function at the singularity is finite and is equal to $g(\phi_{cr})=\frac{2}{\kappa}(a_{cr}-a)$, where we introduced $a_{cr}=\frac{\kappa}{2}(1-\ln\frac{\kappa}{2})$. 
The singularity has a power law that depends on the value of $a$ as we will see in a moment. This is different from the classical case where the curvature singularity was exponentially large. 

Analyzing the solution (\ref{34}) we note that the behaviour of the  metric function $g(\phi)$ has now changed compared to  the classical case.
Now $g(\phi)$ goes to $+\infty$  for $\phi=\infty$ and to $1$ for  $\phi=-\infty$ so that $g(\phi)$ has a minimum at some $\phi=\phi_{min}$. The value $\phi_{min}$ can be easily found 
by solving the condition $g'(\phi_{min})=0$ and is equal to $\phi_{min}=\frac{1}{\kappa}(a-\frac{\kappa}{2})$. The value of $g(\phi)$ at the minimum is
$g(\phi_{min})=1-e^{\frac{2}{\kappa}(a-a_{cr})}$.
One also finds that $\phi_{min}-\phi_{cr}=\frac{1}{\kappa}(a-a_{cr})$.

\begin{figure}
  \includegraphics[width=180mm]{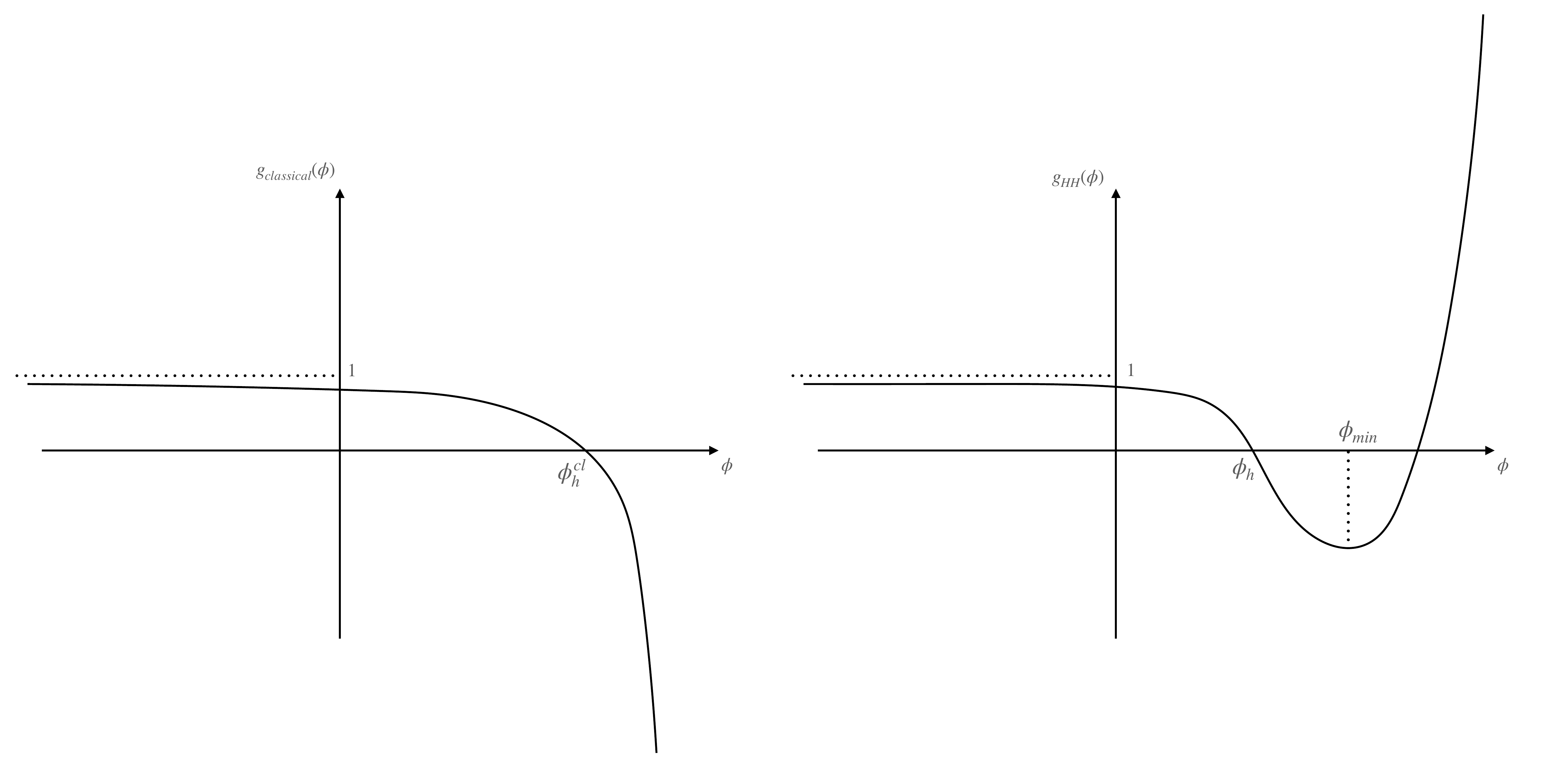}
  \caption{Metric profile for classical vs quantum-corrected (the Harte-Hawking state) black hole}
 \label{fig:classvsHH}
\end{figure}

We can now identify three distinct cases from the discussions above (below we assume that $\kappa>0$):
\medskip

\noindent\underline{i) $a>a_{cr}$:} The metric function $g(\phi)$ is negative at its minimum, $g(\phi_{min})<0$. Hence there exist two values of $\phi$ where $g(\phi)$ vanishes,
$\phi_h$ and $\phi_{h'}> \phi_{h}$.  One also finds that $g(\phi_{cr})<0$ and that $\phi_{min}> \phi_{cr}$. Collecting all inequalities one finds
\be
\phi_h<\phi_{cr}<\phi_{min}<\phi_{h'}\, .
\lb{35}
\ee
At the singularity the curvature behaves as $R\sim \frac{1}{(\phi-\phi_{cr})^3}$.
The shape of the function $g(\phi)$ in this case is shown in figure \ref{fig:classvsHH}.

The branch $\phi>\phi_{cr}$ is completely in the non-classical region. When $\phi$ goes to $+\infty$ the metric becomes  flat  as is seen from (\ref{34-1}).
In this branch the spacetime has a horizon at $\phi=\phi_{h'}$ and a singularity at $\phi_{cr}$.

\medskip

The branch $\phi< \phi_{cr}$  lies in the classical region. The horizon at $\phi=\phi_h$ is a deformation of the classical horizon discussed in section \ref{sec:classdilgrav}.
The Hawking temperature at the quantum-corrected horizon is equal to $T_H=\frac{\lambda}{2\pi}$ as in the classical case. So that the Hawking temperature is not modified.
The same is true for the Hawking temperature at the horizon in the second branch.

\medskip

\noindent\underline{ii) $a=a_{cr}$:}  In this case $g(\phi_{min})=0$ and the two horizon in the first case now merge with the singularity: $\phi_h=\phi_{h'}=\phi_{cr}=\phi_{min}$.
The function $g(\phi)$ then has a double zero that corresponds to an extreme horizon that moreover coincides with the singularity.
The curvature at the singularity now grows  as $R\sim\frac{1}{(\phi-\phi_{cr})}$.

\medskip

\noindent\underline{iii) $a<a_{cr}$:}  The metric function $g(\phi)$ is positive in its minimum and thus is positive everywhere. The curvature singularity is not hidden behind a horizon.
So that this case describes a spacetime with naked singularity.

We note that in the classical case the solutions with horizon and without horizon are separated by value $a=0$ that corresponded to zero mass $M=0$ (see the discussion below \eqref{7}).
In the quantum case the separation of the solutions now happens at $a=a_{cr}$.  Its sign depends on whether value of parameter $\kappa$ is large or small.

\section{Integration of field equations for a general choice  of  quantum state}\label{sec:solvegeneq}
\setcounter{equation}0
\bigskip

Given the discussions in sections \ref{sec:classandpolyakov} and \ref{sec:rst}, a general choice of the quantum state  corresponds to the function $w$,  that appears in (\ref{31}), being non-vanishing. 
We are interested in a static metric
\be
ds^2=-g(x)dt^2+g^{-1}(x)dx^2=-g(\phi)dt^2+g^{-1}(\phi)h^2(\phi) d\phi^2\, ,
\lb{36}
\ee
where in the second equality we choose the dilaton field  $\phi$ as a space-like coordinate and assumed that $\partial_x\phi=h(\phi)$. 
Both the metric function $g(\phi)$ and the function $h(\phi)$ are to be determined from the field equations.
For a static metric the equation $\Box w=0$ can be solved as
\be 
\partial_x w(x)=\frac{C}{g(x)} \, \ {\rm or} \ \ \partial_\phi w(\phi)=\frac{Ch(\phi)}{g(\phi)}\, ,
\lb{37}
\ee
where $C$ is an integration constant. Its value determines the quantum state as was discussed in section \ref{sec:classandpolyakov}.

Assuming that $w$ is a function of the dilaton, (\ref{24}) can be written as (we replace the scalar curvature as $R=-2\Box\phi$)
\be
&&\left(1+\frac{\kappa}{2}(-1+w')\,e^{2\phi}\right)\nabla_\mu\nabla_\nu\phi -\frac{\kappa}{4}\,e^{2\phi}\,\left((-2+w')^2-2w''\right)\nabla_\mu\phi\nabla_\nu\phi \nonumber \\
&&=\frac{1}{2}g_{\mu\nu}\left(\left(1-\frac{\kappa}{2}e^{2\phi}\right)\Box\phi-\frac{\kappa}{4}e^{2\phi}(-2+w')^2(\nabla\phi)^2\right)\, ,
\lb{38}
\ee
where $'$ denotes the derivative with respect to  dilaton $\phi$.
For  the metric  written in $(t, \phi)$ coordinates as in (\ref{36}), one finds
\be
\nabla_t\nabla_t\phi=-\frac{gg'}{2h^2}\, , \ \ \nabla_t\nabla_\phi\phi=0\, , \ \ \nabla_\phi\nabla_\phi\phi=-\frac{g}{2h^2}\left(\frac{h^2}{g}\right)'\, , \ \ \Box\phi=\frac{1}{h}\left(\frac{g}{h}\right)'\, .
\lb{39}
\ee
One then finds that $(t\phi)$ component of equation (\ref{38}) vanishes identically while   the components $(tt)$ and $(\phi\phi)$  lead to the same equation
\be
\left(1-\frac{\kappa}{2}e^{2\phi}\right)\frac{h'}{h}=-\kappa e^{2\phi}\left(\left(1-\frac{Ch}{2g}\right)^2+\frac{C}{2}\frac{g'h}{g^2}\right)\,.
\lb{40}
\ee
Therefore, (\ref{38}) contains only one independent equation.
The other equation that we have to take into account is the dilaton equation (again replacing $R=-2\Box\phi$)
\be
\left(1-\frac{\kappa}{2}e^{2\phi}\right)\Box\phi-2(\nabla\phi)^2+2\lambda^2=0\, .
\lb{41}
\ee
For the metric (\ref{36}) it takes the form
\be
-\left(1-\frac{\kappa}{2}e^{2\phi}\right)\left(\frac{h'}{h}-\frac{g'}{g}\right)-2+2\lambda^2\frac{h^2}{g}=0\, .
\lb{42}
\ee
Equations (\ref{40}) and (\ref{42}) have to be supplemented by the third equation 
\be
R=-2\Box\phi\, ,
\lb{42-1}
\ee
that in the static metric (\ref{36}) (note that $R=-g''(x)$)  takes a simple form
\be
\partial_x^2 g(x)=2\partial_x(g\partial_x\phi)\,.
\lb{43}
\ee
Equivalently, considering $g$  as a function of the dilaton $\phi$, we have
\be
\partial_\phi \left(\frac{1}{h(\phi)} \partial_\phi g(\phi)\right)=2\partial_\phi\left(\frac{1}{h(\phi)}g(\phi)\right)\, .
\lb{44}
\ee
This can be integrated to give
\be
\partial_\phi g(\phi)=2g(\phi)-d \, h(\phi)\, ,
\lb{45}
\ee
where $d$ is an integration constant. Assuming that the solution has the standard asymptotic infinity ($\phi\rightarrow -\infty$), where $\partial_\phi g(\phi)\rightarrow 0$ and the functions in the metric  take values $g(\phi)\rightarrow 1$  and $h\rightarrow -1/\lambda$ as in the classical case,  one  determines the value of the integration constant as
$d=-2\lambda$. 

Equation (\ref{45}) needs to be added to equations (\ref{40}) and (\ref{42}). Since these are three equations on two functions, $g(\phi)$ and $h(\phi)$, one of the equations has to follow from the others.
In fact, as we shall see, the third equation will determine the value of the integration constant that appears when one integrates the first two equations.

First of all  we note that using (\ref{45}),  the equation (\ref{40}) takes the form
\be
\left(1-\frac{\kappa}{2}e^{2\phi}\right)\frac{h'}{h}=-\kappa e^{2\phi}\left(1+\frac{C}{2}\left(\frac{C}{2}-d\right)\frac{h^2}{g^2}\right)\, .
\lb{45-1}
\ee
Also (\ref{45})  can be integrated as follows
\be
g(\phi)=-\frac{d}{2\lambda} \, e^{2\phi} Z(\phi)\, ,
\lb{46}
\ee
where $Z(\phi)$ satisfies the equation
\be
Z'(\phi)=2\lambda e^{-2\phi} h(\phi)\; .
\lb{47}
\ee
Then, the dilaton equation (\ref{42}) can be integrated as
\be
g(\phi)=\frac{2\lambda h(\phi)e^{2\phi}}{\kappa e^{2\phi}-2}( Z(\phi)+A))\, ,
\lb{48}
\ee
where  $A$ is an integration constant to be determined. 
To obtain $Z(\phi)$, we note that using (\ref{46}) to (\ref{48}) we get the following differential equation where $Z(\phi)$ must satisfy
\be
-\frac{d}{2\lambda}(\kappa-2 e^{-2\phi})=Z'(\phi)(1+A/Z(\phi))\, .
\lb{49}
\ee
This equation is easily integrated to give
\be
Z(\phi)+A\ln Z(\phi)=-\frac{d}{2\lambda}(\kappa\phi + e^{-2\phi})+a_1\, ,
\lb{50}
\ee
where $a_1$ is a new integration constant.  Here and in most parts of the paper we consider the domain of positive values $Z>0$. However, we remark
on the case of negative values $Z<0$ in section \ref{subsec:negZ}.
Solving this  equation one gets function  $Z(\phi)$ which can then be used to determine $g(\phi)$ by means of  (\ref{46}), and $h(\phi)$ using (\ref{47}).

Now we substitute  this solution into  equation (\ref{45-1}) and find that (\ref{45-1}) is automatically satisfied provided the integration constant $A$ is related to  constant $C$ in (\ref{37}) as follows
\be
A=-\frac{\kappa}{4d\lambda}C(C-2d)\, .
\lb{51}
\ee
$A$ vanishes for $C=0$ or $C=2d$. In these cases we see that  the resulting solution (\ref{50}), (\ref{46})  is what we had before for the Hartle-Hawking state, where the metric $g_{HH}(\phi)$ takes the form as in (\ref{34}).

Taking $d=-2\lambda$ as explained below \eqref{45},
we find that the complete solution for any constant $C$ is given by 
\be
g(\phi)=e^{2\phi}Z(\phi) \, , \ \ \ h(\phi)=\frac{1}{2\lambda}e^{2\phi}Z'(\phi)\, , \lb{52}
\ee
where $Z(\phi)$ is found by solving the equation
\be
&&Z+A\ln(Z/|A|)=e^{-2\phi}g_{HH}(\phi)\, ,  \lb{52-1} \\
&&g_{HH}(\phi)=1+\kappa\phi e^{2\phi}-a e^{2\phi}\, ,  \  \   A=\frac{\kappa}{8\lambda^2}C(C+4\lambda)\, , \nonumber 
\ee
where $g_{HH}(\phi)$ is the metric function for the Hartle-Hawking quantum state.  Note that we redefined the only remaining integration constant $a_1$
and replaced it by   constant $a$  that appears in the Hartle-Hawking function $g_{HH}(\phi)$.  This constant is eventually related to the mass of the configuration.

Note that for this general solution, the quantum energy density \eqref{13} is given by
\be
T^{(1)0}_{\ \ \ 0}=\frac{\kappa}{2\pi}\left(\frac{6g}{h^2}
+ \frac{4\lambda}{h}
+ \frac{C(C+4\lambda)}{2g}
- \frac{4gh'}{h^3}\right)\,.
\ee
Since $h'$ goes to zero as $\phi$ goes to $-\infty$, the energy density at infinity is given by
\be
T^{(1)0}_{\ \ \ 0\infty} = \frac{\kappa}{4\pi}(C+2\lambda)^{2}
\,, 
\lb{CC}
\ee
similarly to (\ref{16}), so it vanishes for $C=-2\lambda$, which corresponds to the Boulware case. In the presence of a horizon, defined by $g=0$, the energy density is singular except for $C=0$ or $C=-4\lambda$, \textit{i.e} for the Hartle-Hawking state. Therefore the definition of these two quantum states given previously is coherent as we get the same values for the constant $C$.

When $A=0$ (i.e. $C=0$ or $C=2d$) then $g(\phi)=g_{HH}(\phi)$, i.e. the solution (\ref{52}) becomes the quantum-corrected black hole discussed in detail in section \ref{sec:hh}. 
This is the back-reacted geometry for the Hartle-Hawking quantum state.

For non-vanishing $A$, solving the equation in the first line of (\ref{52-1}) one determines the function $Z(\phi)$. Due to the logarithmic term in the equation  one finds that, provided $A\neq 0$, the function   $Z(\phi)$ does not vanish for any finite value of $\phi$.
Since zeros of $Z(\phi)$ determine the zeros of the metric function $g(\phi)$, we conclude that $g(\phi)$ does not have a zero at any finite value of $\phi$. 
Zero of $g(\phi)$ is where the horizon is located.
Thus the geometry (\ref{52}), for $A\neq 0$, is essentially horizon free everywhere in the bulk of the spacetime. The horizon  may, however, appear at the limiting values of $\phi$, either $\phi=+\infty$ or at $\phi=-\infty$, depending on the value of constant $A$. If the scalar curvature is divergent  there, it will indicate that in this case we are dealing with the null singularity.
It appears that the Hartle-Hawking state is the only quantum state for which the back-reacted geometry  contains a regular horizon in the bulk of the spacetime.
In the next sections we shall consider the spacetime (\ref{52}) that arises as a back-reaction of various quantum states and we will provide more concrete examples of the above general statements. 
It should be noted that the analysis of some of the particular cases relevant to the Boulware vacuum present  in the next section has been discussed earlier by Zaslavsky \cite{Zaslavskii:2006pn}.

\section{Back-reacted geometry for the Boulware state}\label{sec:boul}
\setcounter{equation}0
\bigskip
As is discussed in section \ref{sec:classandpolyakov}, the quantum Boulware state corresponds to the value $C=-2\lambda$ and hence we have the corresponding value of constant $A=-\kappa/2$.
The resulting quantum-corrected space-time is described by the metric
\be
&&ds^2=-g(\phi)dt^2+g^{-1}(\phi)h^2(\phi)d\phi^2\, , \nonumber \\
&&g(\phi)=e^{2\phi} Z(\phi)\, ,  \  \  \ h(\phi)=\frac{1}{2\lambda}e^{2\phi}Z'(\phi)\, .
\lb{60}
\ee
The functions $g(\phi)$ and $h(\phi)$ are determined by  function $Z(\phi)$ that is obtained by solving the 
 master equation (\ref{52-1}) that can be presented in the form
\be
&&W(Z)=G(\phi)\, , \nonumber \\
&&  W(Z)=Z-Z_m\ln\frac{Z}{Z_m}\, , \ \  G(\phi)=e^{-2\phi}+\kappa\phi-a\, ,
\lb{61}
\ee
where we define $Z_m=\kappa/2$. 
It is easy to see that function $h(\phi)$ can be also represented as follows
\be
&&h(\phi)=\frac{1}{2\lambda}e^{2\phi}G'(\phi)\frac{Z}{Z-Z_m}\, ,  \nonumber \\
&& \frac{1}{2}e^{2\phi}G'(\phi)=(\frac{\kappa}{2}e^{2\phi}-1)=-F'(\phi)\, ,
\lb{61-1}
\ee
where $F(\phi)$ was earlier defined in (\ref{33}).

On the other hand, $g(\phi)$ is given in implicit form as follows,
\be
e^{-2\phi}=(g-1)^{-1}(Z_m\ln g+a_{cr}-a)\, ,
\lb{61-11}
\ee
where we introduced $a_{cr}=-Z_m\ln Z_m$.

Function $W(Z)$ has its  minimum at $Z=Z_m$, where it takes value $W(Z_m)=Z_m=\kappa/2$. On the other hand, the function $G(\phi)$, provided $\kappa>0$, develops a minimum at $\phi=\phi_{cr}=-\frac{1}{2}\ln\frac{\kappa}{2}$ determined by the condition $G'(\phi_{cr})=0$. Notice that as it is seen  from (\ref{61-1}), generically,  the function $h(\phi_{cr})=0$ at this point.

The scalar curvature computed for the metric (\ref{60})-(\ref{61}) reads
\be
R=\frac{8\lambda^2 e^{-2\phi}}{Z'^3}(ZZ''-Z'^2)\,
\lb{61-0}
\ee
or using (\ref{61-1}), one gets an equivalent form
\be
R=\frac{8\lambda^2 e^{-2\phi}}{ZG'^3}\left(-G'^2 Z+G''(Z-Z_m)^2\right)\, ,
\lb{61-2}
\ee
where all derivatives are taken with respect to $\phi$.
Therefore at  $\phi=\phi_{cr}$ (where $G'(\phi)=0$)  one generically expects  a curvature  singularity. The other point where the curvature appears to be
 singular is where $Z(\phi)=0$.

The value of $G(\phi)$ at the minimum is 
\be
G(\phi_{cr})=W(Z_m)-(a-a_{cr})\, ,
\lb{61-3}
\ee
 where $W(Z_m)=Z_m$.
Note that each function, $W(Z)$ and $G(\phi)$, has two branches: $Z>Z_m$ and $Z<Z_m$ for the function $W(Z)$, and $\phi<\phi_{cr}$ and $\phi>\phi_{cr}$ for the function $G(\phi)$.
The classical domain lies in the quadrant $\phi<\phi_m$, $Z>Z_m$, where $\phi_m$ is defined by condition $G(\phi_m)=Z_m$.

\bigskip

\noindent\underline{Asymptotic behaviour}:  First we analyze the asymptotic behaviour of the metric (\ref{60})-(\ref{61}) in the classical domain, when $\phi\rightarrow -\infty$ and
$Z\rightarrow \infty$. This is where we expect the solution to  approach the classical black hole solution.
Developing the respective asymptotic expansion in equation (\ref{61}) we find that 
\be
&&Z(\phi)=e^{-2\phi}-(a-a_{cr}) +O(\phi e^{2\phi})\, ,\nonumber \\
&&h(\phi)=-1/\lambda +O(\phi e^{2\phi}) \, ,\nonumber \\
&&g(\phi)=1-(a-a_{cr})e^{2\phi}+O(\phi e^{4\phi})\, .
\lb{62}
\ee
We notice the absence of the term $\kappa\phi e^{2\phi}$ in the asymptotic expansion of the metric function $g(\phi)$. As we discussed  earlier in the paper,  this term is due to the 
presence of the thermal radiation at the asymptotic infinity. Since for the Boulware state no such radiation is present, the corresponding term in the metric has to be absent.
This is what we observe in the asymptotic expansion (\ref{62}).

Further analysis of the global structure of the solution depends on the relative position of the minima of the functions $W(Z)$ and $G(\phi)$. There are a total of three cases that we consider below.

\bigskip

\subsection{Global space-time structure: $W(Z_m)>G(\phi_{cr})$ (i.e. $a>a_{cr}$)}

\medskip

\begin{figure}
  \includegraphics[width=180mm]{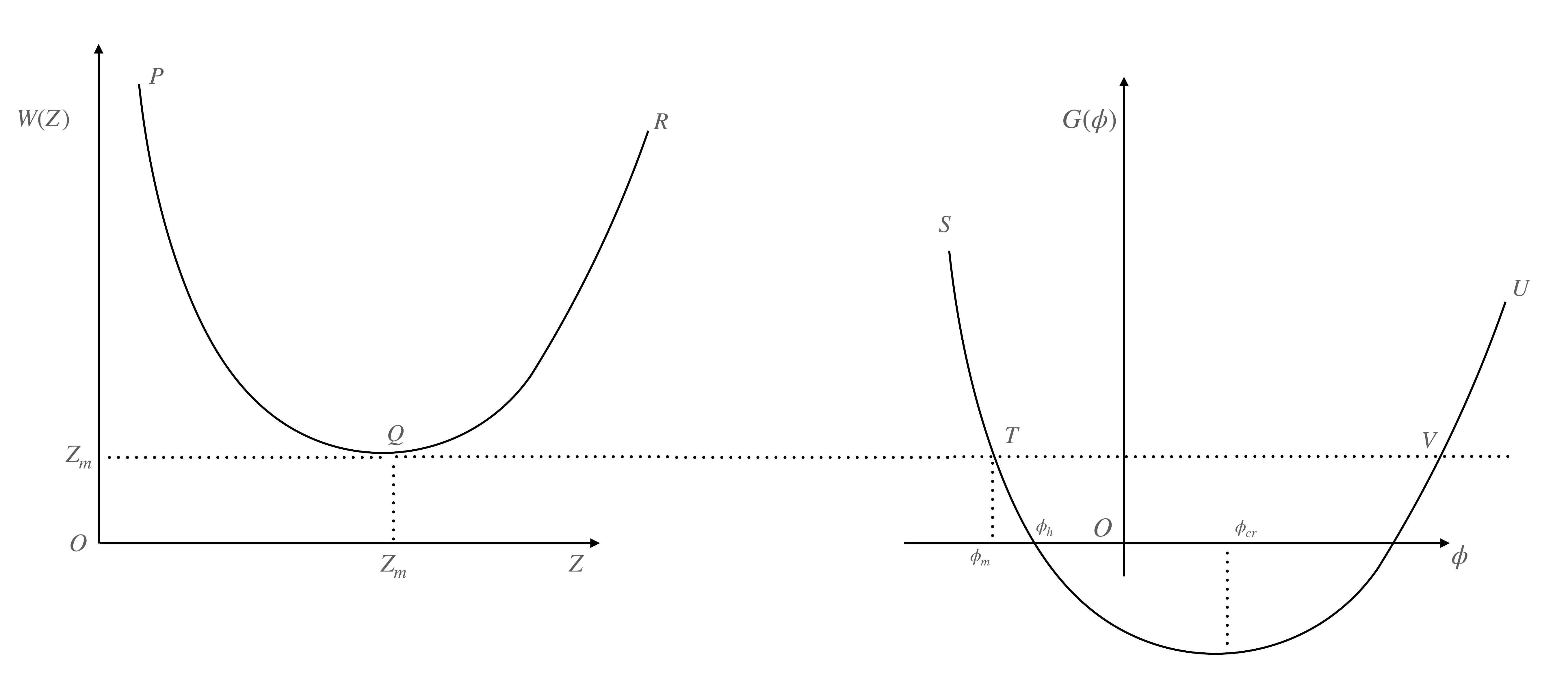}
  \caption{Admissible branches when $W(Z_m)>G(\phi_{cr})$.}
 \label{fig:limit1}
\end{figure}

\noindent\underline{{\it Minimal value of the dilaton}}: In this case, as we start from $\phi=-\infty$, the only branch of function $G(\phi)$  that is accessible  is the ST
branch in  figure \ref{fig:limit1}. It goes for values of $\phi$: $-\infty<\phi\leq \phi_m$, where $\phi_m$ is defined by the condition
$G(\phi_m)=Z_m$.  One finds that $\phi_m=\phi_h(a+\frac{\kappa}{2})$, where  $\phi_h(a)$ is the position of the horizon in the Hartle-Hawking back-reacted geometry for the mass parameter $a$,
$g_{HH}(a, \phi_h)=0$. Thus, for positive $\kappa$ the point $\phi=\phi_m$ is located just outside the horizon in the Hartle-Hawking metric for mass parameter $a$.

\medskip

\noindent\underline{{\it Wormhole interpretation}}: When one reaches the point $\phi=\phi_m$ the value of $\phi$ starts decreasing and it covers the branch ST once again, now in the opposite direction.
On the other hand, values of $Z$ keep decreasing and one goes to the branch $Z<Z_m$ of function $W(Z)$ that goes all the way till $Z=0$.
Thus, the function $e^{-\phi}$, that is similar to the radius $r$ in four dimensions, has a minimum at $\phi=\phi_m$ ($Z=Z_m$). The other way to see this is to compute the gradient of
$\phi$,
\be
(\nabla\phi)^2=\frac{4\lambda^2e^{-2\phi}(Z(\phi)-Z_m)^2}{Z(\phi)\, G'(\phi)^2}\, .
\lb{63}
\ee
It vanishes when $Z=Z_m$. A critical point of $e^{-\phi}$ can be interpreted as a ``minimal surface''. The latter does not have a
good definition in two dimensions so that the condition $(\nabla\phi)^2=0$ is the closest we can get in the analogy with the four-dimensional case.
Thus, in terms of the dilaton $\phi$ we are dealing with a wormhole-type geometry with a throat at  $\phi=\phi_m$.
Note that this does not mean that the metric function $g(\phi)$ has a minimum at $\phi_m$. It already takes a small value at $Z=Z_m$,
\be
g(\phi_m)=\frac{\kappa}{2}e^{2\phi_m}\simeq \frac{\kappa}{2a}=\frac{\kappa}{S_{BH}(a)}\, .
\lb{64}
\ee
In the second equality we considered the limit of large mass, $a\gg \kappa$, and  $S_{BH}=2a$ is the entropy of a classical black hole.
However, $g(\phi)$ keeps decreasing  as soon as one goes to the other branch $Z<Z_m$ of function $W(Z)$.

\medskip

\noindent\underline{{\it Null singularity}}:  One approaches $Z=0$ while $\phi$  goes to $-\infty$. In this regime one solves equation (\ref{61}) which takes the form
\be
Z(\phi)=\frac{\kappa}{2}e^{\frac{2a}{\kappa} }\, e^{-2\phi}\, e^{-\frac{2}{\kappa}e^{-2\phi}}\, .
\lb{65}
\ee
Respectively one finds
\be
g(\phi)=\frac{\kappa}{2}e^{\frac{2a}{\kappa} } e^{-\frac{2}{\kappa}e^{-2\phi}}
\lb{66}
\ee
and
\be
h(\phi)=\frac{1}{\lambda}e^{\frac{2a}{\kappa} } e^{-2\phi}e^{-\frac{2}{\kappa}e^{-2\phi}}\, .
\lb{67}
\ee
\begin{figure}
  \includegraphics[width=180mm]{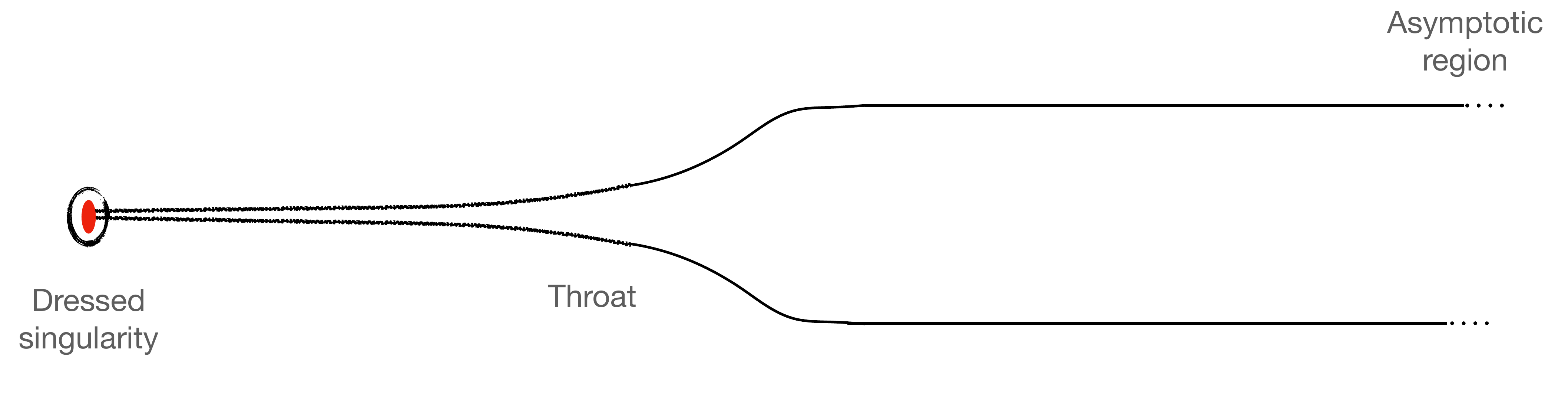}
  \caption{A cross-section of the Euclidean geometry in which the classical horizon is replaced by a `bird's beak' type throat ending with a null singularity.}
 \label{fig:blackthroat}
\end{figure}
The metric function $g(\phi)=-g_{tt}$ goes to zero when $\phi$ goes to $-\infty$ (or $Z(\phi)$ goes to $0$). This indicates the presence of a horizon. 
A bit more careful analysis shows that this new horizon is characterized by the same Hawking temperature $T_H=\frac{\lambda}{2\pi}$ as the classical horizon.
However, this new horizon is singular as the scalar curvature is divergent there
\be
R=2\kappa\lambda^2e^{-\frac{2a}{\kappa}}\, e^{4\phi}\, e^{\frac{2}{\kappa}e^{-2\phi}}\, .
\lb{68}
\ee
So we are dealing with a null singularity. Notice that the position of this singularity is not dependent on the value of the mass parameter $a$.
So that it is not the classical horizon that becomes singular when the back-reaction of the Boulware state  is taken into account.
Instead a new dressed curvature singularity is formed relatively far from the position of the classical horizon. On the other hand, the classical horizon 
is now replaced by a long throat that starts  at $\phi=\phi_m$ ($Z=Z_m$) and continues to shrink till $\phi=-\infty$ $(Z=0)$.
The  Euclidean version of this geometry is illustrated in figure \ref{fig:blackthroat}.

\medskip

\noindent\underline{{\it A long throat picture}}: Let us discuss the long throat picture in somewhat more details.   We call a {\it throat} some region in the space where the metric component $-g_{tt}$ becomes extremely small.  In order to discuss the size of the throat, it is more convenient to go to the optical metric. The metric (\ref{60}) can be rewritten in a conformally flat form,
\be
ds^2=g(Z)\left(-dt^2+d\left(\frac{1}{2\lambda}\ln Z\right)^2\right)\, .
\lb{69}
\ee
The metric in the brackets is the so-called optical metric.  It is the metric in which the rays of light propagate.
 So that the optical distance (or, effectively, the travel time for a light ray)
from a point in space $Z=Z_0> Z_m$ to the point $Z=Z_m=\frac{\kappa}{2}$ is equal to
\be
L=t_H=\frac{1}{2\lambda}\ln Z_0/Z_m\, .
\lb{70}
\ee
This distance becomes large when one takes a small value  for $\kappa$. The point $Z=Z_m$ is thus far away from any other point in the space.
Furthermore, the point $Z_m$ is characterized by a very small value of 
$-g_{tt}$, see (\ref{64}).  All this  justifies the interpretation of $Z_m$ as the neck of the throat. Equation (\ref{70}) gives us an estimate of the size of the throat at $Z_m$.
The throat however does not stop at $Z_m$. It continues further for $Z<Z_m$ till $Z=0$ where it is ending with a null singularity. 
The optical ``size'' of the extended throat
 is  infinite.

\bigskip

\subsection{ Global space-time structure: $W(Z_m)=G(\phi_{cr})$ (i.e. $a=a_{cr}$)}

\medskip

\begin{figure}
  \includegraphics[width=180mm]{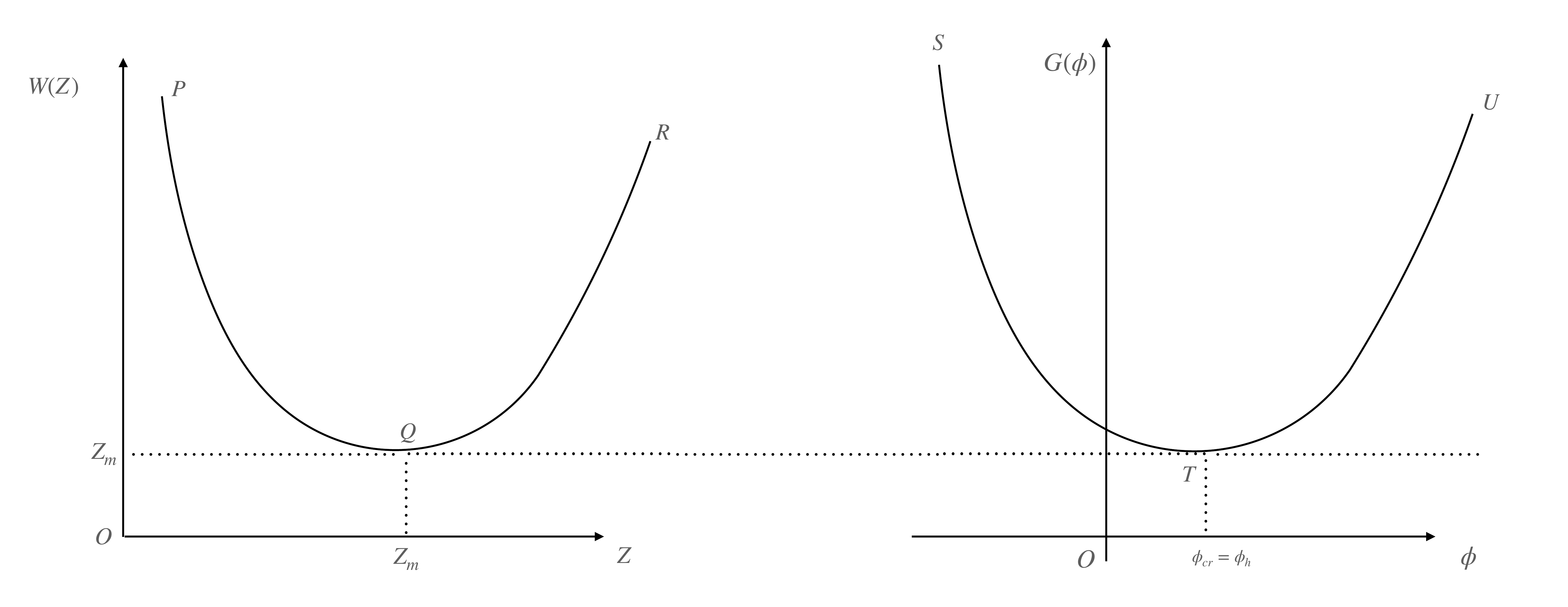}
  \caption{Admissible branches when $W(Z_m)=G(\phi_{cr})$.}
 \label{fig:limit_same}
\end{figure}

The two minima of the functions in (\ref{61}) coincide in this case (see figure  \ref{fig:limit_same}).
The master equation (\ref{61}) then  takes the form that is more convenient to express in terms of a new variable $y=e^{-2\phi}$,
\be
W_0(Z)=W_0(y)\, , \  \  \  W_0(Z)=Z-Z_m\ln Z\, .
\lb{71}
\ee
The function $W_0(Z)$ has two branches: $Z>Z_m$ and $Z<Z_m$. Therefore, there are two  solutions to this equation. If $Z$ and $y$ are from the 
same branch, i.e. $Z(y)>Z_m$ while $y> Z_m$ or $Z(y)<Z_m$ while $y<Z_m$, we call this solution {\it direct}. If  $Z(y)$ and $y$ are from different branches,
i.e. $Z(y)<Z_m$ if $y> Z_m$ and $Z(y)> Z_m$ if $y< Z_m$ we call this solution {\it twisted}. We start our analysis with the simplest one.

\medskip

\noindent\underline{{\it Direct solution}}:
The direct solution to (\ref{71}) is very simple,
\be
Z(y)=y=e^{-2\phi}\, .
\lb{72}
\ee
Respectively we find that
\be
g(\phi)=1\, ,  \  \  {\rm and } \  \   h(\phi)=-1/\lambda \, .
\lb{73}
\ee
This is the classical linear dilaton solution. The spacetime is Minkowski. It appears as a Boulware type solution in the 
two-dimensional semiclassical RST model.

\medskip

\noindent\underline{{\it Twisted solution}}:
As we explained above, in the twisted solution $Z(y)$ and $y$ belong to different branches. We did not find an explicit analytic form of $Z(y)$ in this case.
However, it can be easily found in certain limits. 

First of all we consider the equation (\ref{71}) near the minimum $Z=Z_m$, where the function $W(Z)$ expands as follows
$W(Z)=W(Z_m)+\frac{1}{2Z_m}(Z-Z_m)^2-\frac{1}{3Z_m^2}(Z-Z_m)^3+\dots$. Therefore equation (\ref{71}) can be solved as
\be
e^{-2\phi}=y=Z_m+(Z_m-Z)+\frac{2}{3Z_m}(Z_m-Z)^2-\frac{4}{9Z_m^2}(Z-Z_m)^3+\dots \, .
\lb{74}
\ee
As a result, we find that
\be
&&g(\phi)=1-\frac{2}{Z_m}(Z_m-Z)+\frac{4}{3Z_m^2}(Z_m-Z)^2+\dots\, , \nonumber \\
&&h(\phi)=\frac{1}{\lambda}\left(1+\frac{4}{3Z_m}(Z-Z_m)+\frac{4}{9Z_m^2}(Z-Z_m)^2+\dots \right) \, .
\lb{75}
\ee
The point $\phi(Z=Z_m)=\phi_{cr}$ is the critical point at which the singularity in the scalar curvature may appear since $G'(\phi)$ vanishes at this point.  
Indeed, we find that $G'=-2(Z_m-Z)$, where we keep only the leading terms. 
However, the function that appears in the numerator in (\ref{61-2})  also has a simple root at $Z=Z_m$, $(Z-Z_m)^2G''-ZG'^2=-8(Z-Z_m)^3$.
So that the zeros in the numerator and in the denominator mutually cancel and the curvature comes out regular,
$R=-8\lambda^2$.

We note that by exactly the same mechanism the gradient of the dilaton does not vanish at $Z=Z_m$ and is equal to $(\nabla\phi)^2=\lambda^2$. 
So that in this case there is no wormhole type behaviour for the dilaton. Also, since $g=1$ at $Z=Z_m$, there is no throat there. 

Next, we consider the limit when $\phi$ goes  to $-\infty$, i.e. $y\rightarrow +\infty$,  while $Z(y)\rightarrow 0$. The equation (\ref{71}) in this limit solves as 
\be
Z(y)=e^{-{y}/{Z_m}}\,, \  \  y=e^{-2\phi}\, .
\lb{76}
\ee
Once again, in order to simplify the formulas we express everything in terms of variable $y=e^{-2\phi}$, which gives the metric functions as
\be
g(\phi)=y^{-1}e^{-y/Z_m}\, , \ \  h(\phi)=\frac{1}{\lambda Z_m}e^{-y/Z_m}\, .
\lb{77}
\ee
The scalar curvature is found to be
\be
R=-4\lambda^2Z^2_m y^{-1}e^{y/Z_m}\, ,
\lb{78}
\ee
which is divergent in this limit. We thus have a curvature singularity. Note, that this behaviour is drastically different from the 
classical one where in this limit the spacetime becomes flat. Notice that the singularity is again null since  $g(\phi)$ is  vanishing in this limit.

Finally, we consider the limit when $\phi\rightarrow +\infty$ ($y\rightarrow 0$) while $Z(y)\rightarrow +\infty$.
We find, keeping only the leading terms,
\be
Z(\phi)=2Z_m\phi\, , \ \  g(\phi)=2Z_m\phi e^{2\phi}\, , \ \ h(\phi)=\frac{Z_m}{\lambda}e^{2\phi}\, .
\lb{79}
\ee
The scalar curvature in the considered limit
\be
R=-\frac{4\lambda^2}{Z_m}e^{-2\phi}\, 
\lb{80}
\ee
now goes to zero. Thus, the spacetime is asymptotically flat in this limit. 

\medskip

We conclude that the {\it twisted} solution describes a spacetime which is asymptotically flat at one end and has a null singularity at the other.
We note also that  in any limits it does not approach the classical solution.

\bigskip

\subsection{Global space-time structure: $W(Z_m)<G(\phi_{cr})$ (i.e. $a<a_{cr}$)}\label{subsec:7.3}

\medskip

\begin{figure}
  \includegraphics[width=180mm]{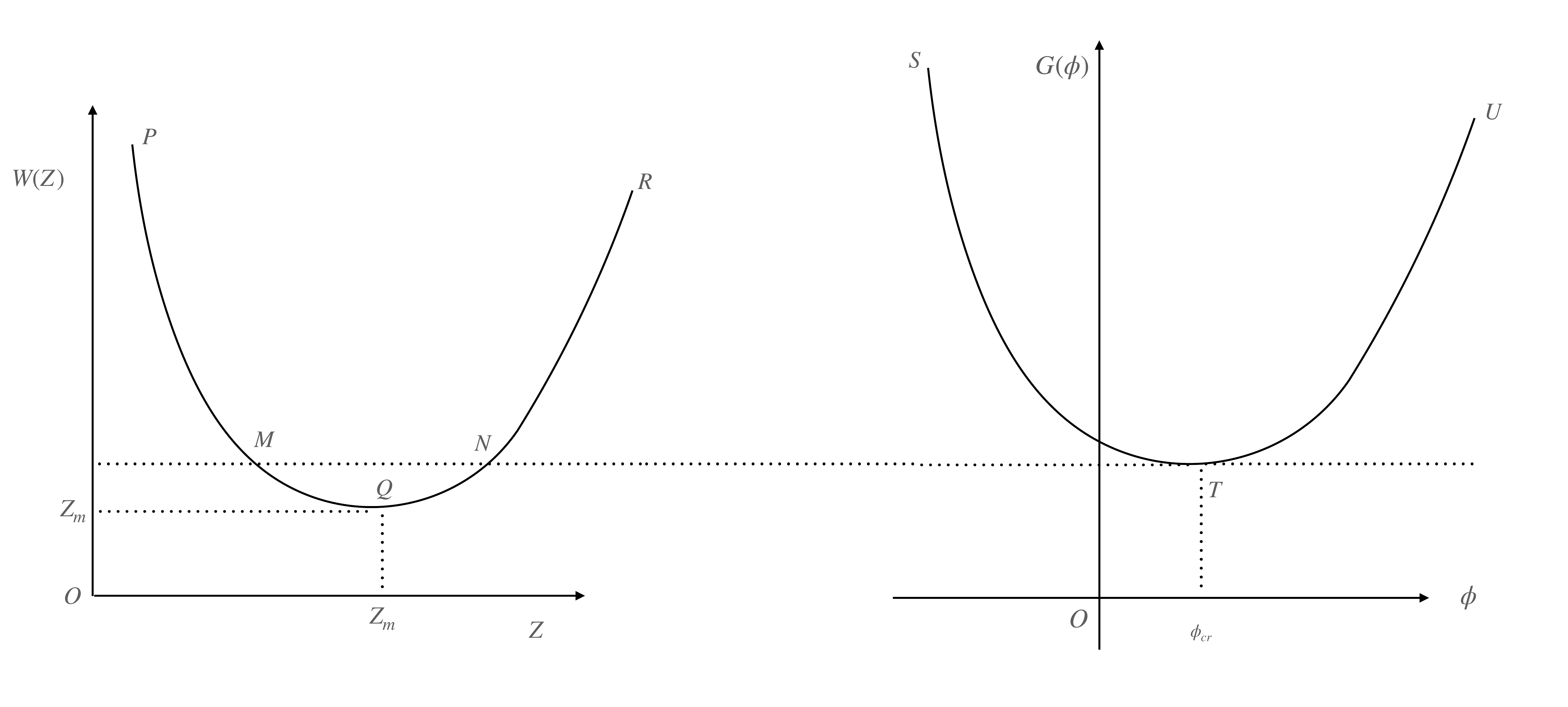}
  \caption{Both $PM$ and $NR$ branches are admissible for $W(Z_m)<G(\phi_{cr})$.}
 \label{fig:boulcase3}
\end{figure}

In this case the minimum of function $W(Z_m)$ is lower than the minimum of function $G(\phi)$ (see e.g. figure \ref{fig:boulcase3}).  Since one expects a curvature singularity to appear at $\phi=\phi_{cr}$, the possible values
for $\phi$ are either $\phi<\phi_{cr}$ (classical branch) or $\phi>\phi_{cr}$ (non-classical branch).  Solving equation (\ref{61}) at $\phi=\phi_{cr}$ one finds two solutions:
$Z_1$ and $Z_2$ such that $Z_2<Z_m<Z_1$. Therefore there are two accessible branches: $Z>Z_1$ and $Z<Z_2$ for $Z$.
For simplicity below we shall analyze only the classical branch $\phi<\phi_{cr}$. Since this can be combined with two possible branches for $Z$ we have two possible solutions, which by 
analogy with the case $a=a_{cr}$, we shall call {\it direct} and {\it twisted} solutions.

\medskip

\noindent\underline{{\it Direct solution: $\phi<\phi_{cr}$ and $Z>Z_1$.}}
The asymptotic behaviour for $\phi\rightarrow -\infty$ and $Z\rightarrow +\infty$ was already analyzed above, see (\ref{62}). 
Therefore the spacetime is asymptotically flat. However, as expected, the curvature becomes divergent when one approaches value $\phi=\phi_{cr}$,
\be
R=\frac{\lambda^2}{2}\frac{(Z_1-Z_m)^2}{Z_1Z_m}(\phi-\phi_{cr})^{-3}\, .
\lb{81}
\ee
There is also no horizon at this limit, so we conclude that the {\it direct} solution describes a naked singularity. We also note that since $a<a_{cr}$, the second term in the 
asymptotic expansion  $g(\phi)=1-(a-a_{cr})e^{2\phi}$ is negative. Therefore this is similar to the negative mass case in the classical black hole solution.

\medskip

\noindent\underline{{\it Twisted solution: $\phi<\phi_{cr}$ and $0<Z<Z_2$.}}  This solution also has a singularity at $\phi=\phi_{cr}$ ($Z=Z_2$). The curvature there is described by the same formula
(\ref{81}) by replacing $Z_1$ with $Z_2$. On the other end, when $\phi\rightarrow -\infty$ one finds the asymptotic expansion (once again using the variable $y=e^{-2\phi}$ in order to simplify the expressions)
\be
Z(\phi)=Z_m e^{-y/Z_m}\, , \ \  g(\phi)=Z_m y^{-1} e^{-y/Z_m}\, , \  \  h(\phi)=\frac{1}{\lambda} e^{-y/Z_m}\,,
\lb{82}
\ee
and the scalar curvature takes the form
\be
R=-4\lambda^2 Z_m y^{-1}e^{y/Z_m}\, ,
\lb{83}
\ee
which indicates the presence of the singularity. Since $g(\phi)$ in (\ref{82}) goes to zero in this limit it is a null singularity.
Thus, the {\it twisted} solution represents a spacetime with a naked singularity at one end and a null singularity at the other.

\subsection{The case $Z<0$}\label{subsec:negZ}
\medskip
As mentioned earlier, so far we have only considered the branch where $Z$ takes positive values. In order to get a complete picture, we can consider what happens when $Z$ evolves on the branch with negative values. On this branch, (\ref{61}) becomes
\be
&&W(Z)=G(\phi)\, , \ \ Z<0 \nonumber \\
&&  W(Z)=Z-Z_m\ln\frac{-Z}{Z_m}\, , \ \  G(\phi)=e^{-2\phi}+\kappa\phi-a\, ,
\lb{Z1}
\ee
where $W(Z)$ is a monotonic function with limits given by
$ W(Z) {\rightarrow} - \infty$ (when $Z\rightarrow -\infty$)  and $W(Z) {\rightarrow} +\infty$ (when $Z\rightarrow 0$).
Therefore, the spacetime is in the region $Z> Z_{cr}$, where $Z_{cr}$ is the solution of $W(Z_{cr}) = G(\phi_{cr})$. The spatial infinity $\phi \rightarrow -\infty$ corresponds to $Z\rightarrow 0$, that is to say to the null singularity discussed previously. When $\phi$ approaches $\phi_{cr}$, we arrive at the curvature singularity, and $Z$ never vanishes so that there is no horizon. Therefore the branch $Z<0$ corresponds to a spacetime located between a null singularity at infinity and a naked singularity.

\subsection{ The case $\kappa=-k<0$}\label{subsec:negkappa}

In this section we briefly discuss the case of negative values of $\kappa$. 
Indeed, if we consider non-physical fields, e.g. {\it ghosts}, then it will contribute negatively to the 2d central charge, i.e. to the coupling $\kappa=(N-24)/24$.
If these non-physical fields dominate then $\kappa$ can be negative. This is particularly the case if e.g. there are no physical fields at all, and $\kappa$ is induced only by quantum dilaton gravity.
Therefore, for the sake of completeness, it is worth considering it in some details (we will come back to a similar set-up in subsection \ref{subsec:genCnegkappa} and in section \ref{sec:hybrid} later).  Note that since all fields, including the non-physical ones, are in the Boulware state, they are not visible at the asymptotic infinity. The master equation in this case is
\be
&&W(Z)=G(\phi)\, , \nonumber \\
&&W(Z)=Z+Z_m\ln\frac{Z}{Z_m}\, ,  \  \  G(\phi)=e^{-2\phi}-k\phi-a\, ,
\lb{Z3}
\ee
where now $Z_m=k/2>0$. This case is interesting since both functions $W(Z)$ and $G(\phi)$ are monotonic (note that we consider only the region $Z>0$). 
$W(Z)$ is monotonically increasing while $G(\phi)$ is monotonically decreasing. Therefore, the solution explores all possible values between  $ -\infty<\phi<+\infty$ and $0<Z<+\infty$.
In the limit $\phi\rightarrow -\infty$ the solution is asymptotically Minkowski as we showed in the beginning of section \ref{sec:boul}. The asymptotic value of the metric function is $g=1$.
On the other end of the spacetime, when $\phi\rightarrow +\infty$, perturbatively solving  equation (\ref{Z3}) one finds
\be
Z(\phi)=Z_0(\phi)e^{-Z_0(\phi)/Z_m}\, ,  \  \   Z_0(\phi)=Z_m e^{-a/Z_m}e^{-2\phi}\, .
\lb{Z4}
\ee
Notice that $Z_0(\phi)\rightarrow 0$ when $\phi\rightarrow +\infty$.
Therefore one finds that
\be
g(\phi)=Z_me^{-a/Z_m} \, e^{-Z_0(\phi)/Z_m}\, , \  \   h(\phi)=-\lambda^{-1}Z_m e^{-a/Z_m}e^{-Z_0(\phi)/Z_m}\,. 
\lb{Z5}
\ee
When $\phi=+\infty$, they have the limiting values
\be
g(\phi=+\infty)=Z_m e^{-a/Z_m}\, ,  \  \ h(\phi=+\infty)=-\lambda^{-1}Z_m e^{-a/Z_m}\, .
\lb{Z6}
\ee
The scalar curvature in this limit
\be
R=\frac{4\lambda^2}{Z_m}e^{-2\phi}
\lb{Z7}
\ee
approaches zero. So that at this end the spacetime is again asymptotically flat.  

Thus the case of negative $\kappa=-k$ is interesting since in this case the spacetime solution is everywhere regular, which at one end $(\phi=-\infty)$ 
approaches the classical black hole metric and at the other end $(\phi=+\infty)$ is again asymptotically flat. One can show that provided the mass parameter
$a>Z_m\ln Z_m$, the metric function is monotonically decreasing $g'(\phi)<0$. It goes from $g=1$ at one end to its minimal value
\be
{\rm min}\,  g=g(\phi=+\infty)=\frac{k}{2}e^{-2a/k}
\lb{Z8}
\ee
at the other end. Since the classical black hole entropy $S_{BH}=2a$, this minimal value is exponentially small (for large $a$)  in terms of the classical entropy,
\be
{\rm min}\,  (-g_{tt})=\frac{k}{2}e^{-S_{BH}/k}\, .
\lb{Z9}
\ee
This is similar to the bound found in four dimensions in  \cite{Berthiere:2017tms}.
The asymptotic region of large, positive value of the dilaton can thus be represented as a very long throat in which the metric function is extremely small while non-zero.
This is an example of a horizonless geometry which is regular everywhere and acts as a black hole mimicker.

\section{Back-reacted geometry for a general $C$-state}\label{sec:genC}
\setcounter{equation}0
\bigskip

After considering the Hartle-Hawking (HH) and the Boulware states in detail, we now turn our attention to the case of arbitrary $C$. The master equations that we should start with are then
 \eqref{52} and \eqref{52-1}. Any value of $C\neq -2\lambda$ (otherwise the state is Boulware) and $C\neq0,-4\lambda$ (otherwise the state is HH) falls in this case. These other values of $C$ effectively determines the sign of $A$, which in turn appears in the master equations.

\medskip

\noindent\underline{\it Asymptotic behaviour:}  In the classical domain, where $\phi\rightarrow -\infty$ and $Z\rightarrow +\infty$, equations   \eqref{52} and \eqref{52-1}
can be solved asymptotically. One finds that
\be
&&Z=e^{-2\phi}+ \frac{\kappa}{4\lambda^2}(C+2\lambda)^2\phi\, , \nonumber \\
&&g=1+    \frac{\kappa}{4\lambda^2}(C+2\lambda)^2 \phi e^{2\phi}\, , \nonumber \\
&&h=-\frac{1}{\lambda}+  \frac{\kappa}{8\lambda^2}(C+2\lambda)^2 e^{2\phi}\, .
\lb{S-1}
\ee
This is consistent with the asymptotic perturbation (\ref{b-6}), (\ref{b-7})  of the metric over the linear vacuum produced by the thermal Hawking radiation with the energy density  \eqref{16}.

\subsection{The case $A<0$}

 In fact, if $A=\frac{\kappa}{8\lambda^2}C(C+4\lambda)<0$, then we have 
\be\label{eq:genAneg-1}
&&W(Z)=G(\phi)\, \qquad\text{with} \nonumber \\
&&  W(Z)=Z-|A|\ln\frac{Z}{|A|}\, \qquad\text{and} \qquad  G(\phi)=e^{-2\phi}+\kappa\phi-a\, .
\ee
This is effectively the Boulware case considered in \eqref{61} of the above section. The only difference now being that $Z_m$ needs to be substituted by $|A|$. Hence all the sub-cases exactly follow the various cases we considered in section \ref{sec:boul}, and hence we will not be discussing this scenario in any further details.

\subsection{The case of $A>0$}

However, the situation is quite different when we have $A>0$. In this case, we can once again define our master equation by 
\be\label{eq:genAneg}
&&W(Z)=G(\phi)\, \qquad\text{with} \nonumber \\
&&  W(Z)=Z+A\ln\frac{Z}{A}\, \qquad\text{and} \qquad  G(\phi)=e^{-2\phi}+\kappa\phi-a\, ,
\ee
along with 
\be\label{eq:genCmet}
g(\phi)=e^{2\phi} Z(\phi)\, ,  \  \  \ h(\phi)=\frac{1}{2\lambda}e^{2\phi}Z'(\phi)\, .
\ee
The first thing we notice is that the behavior of $W(Z)$ is monotonic in this case and is given by the left plot of figure \ref{fig:genCcase1}. This immediately lets us conclude that the value of the zero of the $W(Z)$ is located at $Z=Z_0\approx 0.567 A \neq 0$, which corresponds to $\phi=\phi_h$ (the location of the horizon in the Hartle-Hawking solution). Therefore, due to quantum modifications in these general $C$-states and for $A>0$, we have a throat of finite size at the location of the HH horizon. At this location, the metric function $-g_{tt}$ takes a very small non-zero value
\begin{equation}
	g(\phi)\to e^{2\phi_h} Z_0\simeq \frac{Z_0}{S_{BH}}\,.
\end{equation} 

We can also compute the general expression of Ricci scalar for this geometry, which is given precisely by \eqref{61-0} above,
which for this particular case of \eqref{eq:genAneg} becomes
\be\label{eq:genCR}
R=\frac{8\lambda^2 e^{-2\phi}}{ZG'^3}\left(-G'^2 Z+G''(Z+A)^2\right)\,.
\ee

Note that, as long as $\kappa>0$,  the behavior of $G(\phi)$ still remains the same  (we will shortly discuss the $\kappa<0$ case in this context in the next subsection). In particular, it still has a minima at $\phi=\phi_{cr}=-\frac{1}{2}\ln\frac{\kappa}{2}$ determined by the condition $G'(\phi_{cr})=0$. Its value at the minima is again given by 
\begin{equation}
	G(\phi_{cr})=a_{cr}-a\qquad\text{with}\qquad a_{cr}=\frac{\kappa}{2}\left(1-\ln\frac{\kappa}{2}\right)\,.
\end{equation}
This definition of $a_{cr}$ is slightly different from what we've defined in \eqref{61-3}, and is rather similar to our definition below \eqref{34-1}.
However, we can end-up with various comparative situations depending on the sign of $G(\phi_{cr})$. These are the cases we will enumerate next. 

\bigskip

\subsubsection{The cases $G(\phi_{cr})<0$ (or $a>a_{cr}$) and $G(\phi_{cr})\geq 0$ (or $a_{cr}\geq a$)}

\medskip
\begin{figure}
  \includegraphics[width=180mm]{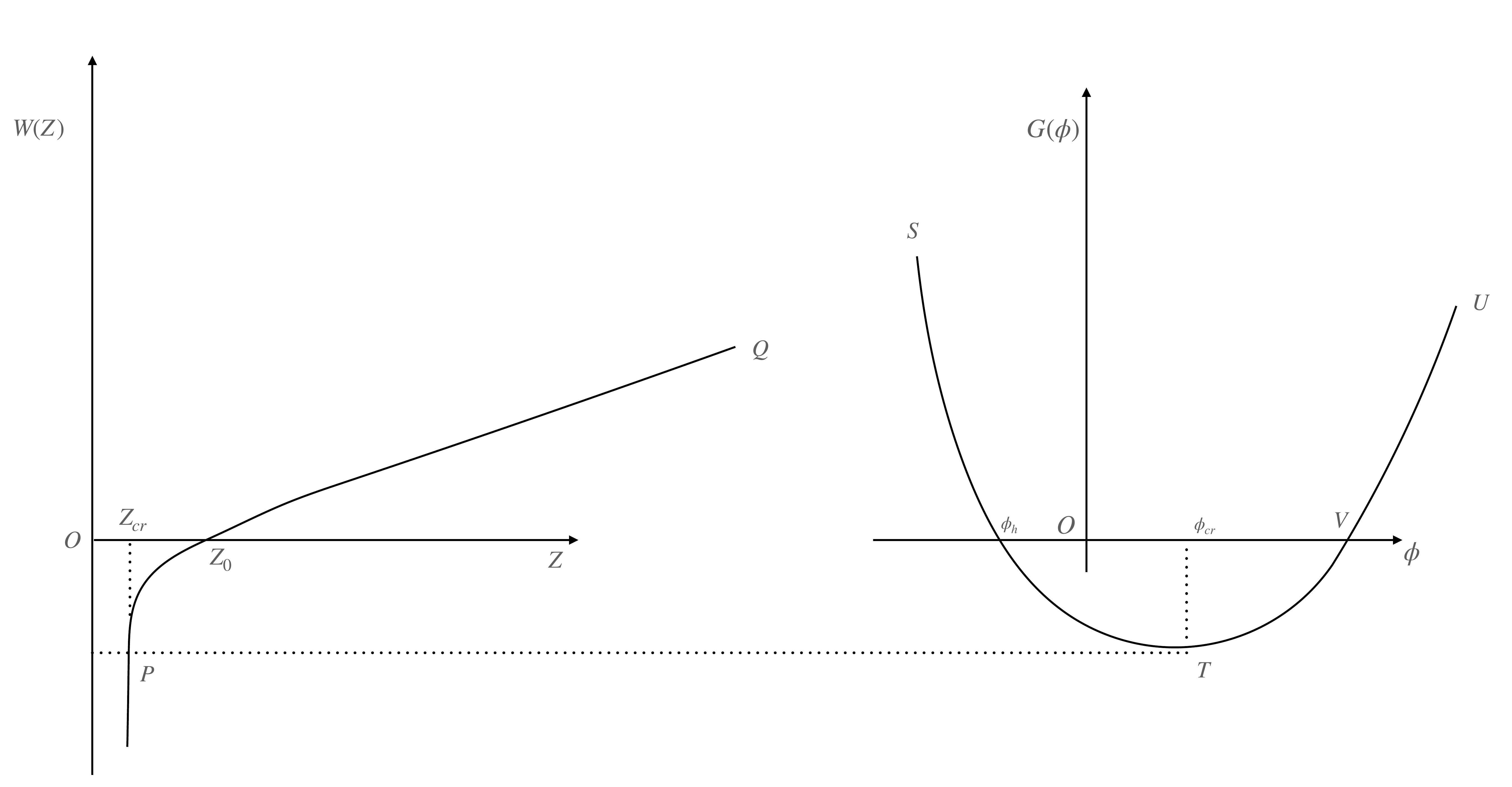}
  \caption{The $PQ$ branch can correspond to either $ST$ or $TVU$ branch for $G(\phi_{cr})>0$.}
\label{fig:genCcase1}
\end{figure}

As we will see below, both these cases yield the same geometric structure of the spacetime. These correspond to the situation we have in figure \ref{fig:genCcase1} (in this figure we have plotted the $G(\phi_{cr})<0$ case, but a similar picture is there for $G(\phi_{cr})\geq 0$). The minimum of $G(\phi)$ at $\phi=\phi_{cr}$ correspond to a value of $Z=Z_{cr}$. As at this point $G'(\phi)=0$, the resulting spacetime has a curvature singularity coming from the divergent piece of \eqref{eq:genCR}. This point corresponds to a naked singularity as the metric function \eqref{eq:genCmet} is still finite at this point.

\medskip

\noindent\underline{i) $Z>Z_{cr}$ and $\phi<\phi_{cr}$:} As we go from $P$ to $Q$ in the plot of $W(Z)$ in figure \ref{fig:genCcase1} above, we can either choose the $TS$ or the $TVU$ branch for $G(\phi)$. The $TS$ branch once again corresponds to the classical case designated by $\phi\to -\infty$ and $Z\to \infty$ and it was already studied before in \eqref{S-1}. Therefore in this branch, starting from asymptotically flat spacetimes (with subleading corrections due to thermal radiation), we get to a naked singularity after passing through a throat at $\phi=\phi_h$.

\medskip

\noindent\underline{ii) $Z>Z_{cr}$ and $\phi>\phi_{cr}$:}
On the other hand, the branch $TVU$ entirely denotes a quantum spacetime. Asymptotically at $U$, $\phi\to \infty$ and $Z\to \infty$. In this limit, the analysis follows the same steps as \eqref{79} and \eqref{80}. At the end, we have an asymptotically flat solution with the Ricci scalar (note that it doesn't depend on $A$)
\begin{equation}
	R=-\frac{8\lambda^2}{\kappa}e^{-2\phi}\,.
\end{equation}
On the other hand, at point $V$ the metric function satisfies the classical HH condition of $g_{HH}=0$. However, once again it doesn't correspond to the metric function being zero.

The analysis is essentially same also for $G(\phi_{cr})> 0$. So we see that both these cases produce a spacetime which has a naked singularity at one end and an asymptotically flat spacetime on the other.

\subsection{The case of negative $\kappa$}\label{subsec:genCnegkappa}

\begin{figure}
  \includegraphics[width=180mm]{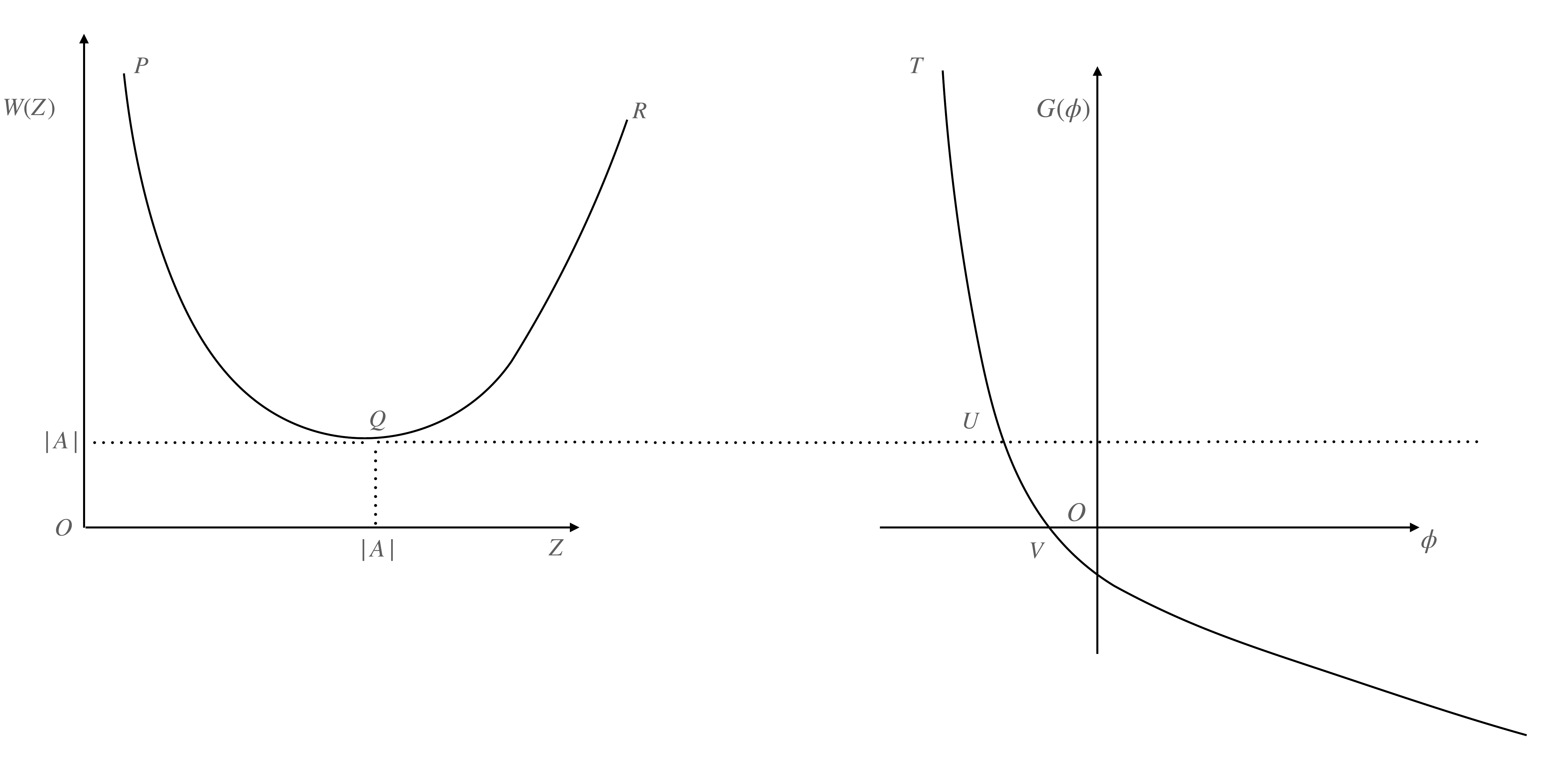}
  \caption{The case of negative $A$ and $\kappa$. As we go from $T$ (at $\phi\to -\infty$) to $U$, we traverse the entire $W(Z)$.}
 \label{fig:negkgenC}
\end{figure}

Following our steps in subsection \ref{subsec:negkappa}, we briefly discuss here the case of negative $\kappa=-k<0$. The resulting spacetime once again falls into the scenarios we encountered before. In particular, if $A>0$, then the corresponding master equations are just what we have in \eqref{Z3} with $Z_m$ now replaced by $A$. So in this case, we once again end up with a black hole mimicker geometry with a throat of exponentially small (in its classical entropy) size.

The situation is slightly different when $A<0$. As now we have a minima for the function $W(Z)$, but the function $G(\phi)$ is monotonic. This situation is illustrated in figure \ref{fig:negkgenC} (we have only plotted and studied the case when $Z>0$. Similar arguments can be made for $Z<0$ following the arguments of subsection \ref{subsec:negZ}). When we approach $\phi\to -\infty$, we can either approach $P$ or $R$. If we approach the point $R$ asymptotically, then we are in the classical regime, with the spacetime approaching asymptotically flat spacetimes following the steps around \eqref{S-1}. On the other hand, when we approach $P$ ($Z\to 0$), we have
\begin{equation}
	Z=|A|\, e^{-y/|A|}\qquad\text{with}\qquad y=e^{-2\phi}\,,
\end{equation}
which yields the metric functions to be 
\begin{equation}
	g(y)=\frac{|A|}{y}\, e^{-\frac{y}{|A|}}\quad\text{and}\quad h(y)=\frac{1}{\lambda}\, e^{-y/|A|}\,.
\end{equation}
The resulting spacetime is once again null singular as we approach $y\to \infty$. So, in this case, we have an asymptotically flat spacetime on one end, and a null singularity on the other.

\section{Hybrid  quantum state}\label{sec:hybrid}
\setcounter{equation}0
\bigskip
In this section we would like to explore one more interesting possibility. It is possible that among the quantum fields, some of them (characterized by coupling $\kappa_1$) are in the Hartle-Hawking state and the others (with the coupling $\kappa_2$) are in the Boulware state.  This situation appears quite naturally when besides the physical quantum fields  (with positive $\kappa_1>0$)
we also have the unphysical fields, or ghosts, with negative $\kappa_2<0$. We do not expect to see a thermal radiation made of ghosts at the asymptotic infinity so that the ghost fields 
have to be always in the Boulware state. This is  the situation when the Boulware state is the distinguished and, in fact, the only possible choice for the quantum state.
The physical fields, on the other hand,  may be in the Hartle-Hawking state. So that any observer at the asymptotic infinity would see only the thermal radiation made of the physical particles.
This situation will be our main focus in the discussion below.

In the present situation it is natural to introduce two auxiliary fields, $\psi_1$ and $\psi_2$, so that the Polyakov action will be a sum
\be
I_1=I_1^{(1)}+I_1^{(2)}\, , \ \  I_1^{(n)}=-\frac{\kappa_n}{2\pi}\int_M d^2 x\sqrt{-g}\,\left(\frac{1}{2}(\nabla\psi_n)^2+\psi_n R\right)\, , \ \ n=1,\, 2\,.
\lb{100}
\ee
The action of the RST model then  is the same as before,
\be
I_{RST}=I_0+I_1+I_2\, , \ \  I_2=-\frac{(\kappa_1+\kappa_2)}{2\pi}\int d^2 x \sqrt{-g}\phi R\,.
\lb{101}
\ee
On the level of the action and the field equations this separation is of course redundant  and one always can return to one auxiliary field with one coupling $\kappa=\kappa_1+\kappa_2$.
The important difference, however, appears when one makes a choice of the quantum state. One has that
\be
\psi_n=-2\phi+w_n\, , \  \  \Box w_n=0\, , \  \  w'_n (\phi)=\frac{C_n h(\phi)}{g(\phi)}\, , \  n=1,2
\lb{102}
\ee
where $C_1\neq C_2$.
The integration of the field equations in this more general situation goes through in the same way as before and one arrives at the equation  (\ref{52-1})
which now takes the form
\be
&&Z+A\ln(Z/|A|)=e^{-2\phi}g_{HH}(\phi)\, ,  \ g_{HH}(\phi)=1+(\kappa_1+\kappa_2)\phi e^{2\phi}-a e^{2\phi}\, ,  \lb{103} \\
&&A=A_1+A_2\, , \ A_n=\frac{\kappa_n}{8\lambda^2}C_n(C_n+4\lambda)\, , \ n=1,2\nonumber 
\ee

We will choose different quantum states for $\psi_1$ and $\psi_2$. Namely, $\psi_1$ will correspond to the Hartle-Hawking state, i.e. $C_1=0$.
On the other hand, $\psi_2$ will be in the Boulware state, $C_2=-2\lambda$.  The general situation (arbitrary $\kappa_1$ and $\kappa_2$)
can be easily analyzed.  For simplicity and for the purposes of the illustration of how this new situation is different from the case when all fields are in the same 
quantum state, we shall consider the case when the total $\kappa$ vanishes i.e. $\kappa=\kappa_1+\kappa_2=0$.  In other words, the contribution of ghosts
($\kappa_2<0$) is precisely compensated by the contribution of the  physical fields $(\kappa_1=-\kappa_2>0$). For this choice of the quantum states
one finds $A_1=0$ and $A=A_2=-\frac{\kappa_2}{2}=\frac{\kappa_1}{2}>0$. The master equation (\ref{103}) then takes the form
\be
Z+Z_m\ln Z/Z_m=e^{-2\phi}g_{cl}(\phi)\, , \  g_{cl}(\phi)=1-ae^{2\phi}\, , \ Z_m=\frac{\kappa_1}{2}>0\, .
\lb{104}
\ee
Notice that $\phi e^{2\phi}$ term in the metric function $g_{HH}(\phi)$ disappears and it becomes the metric function that appears in the classical black hole solution
(\ref{7}).

Asymptotically, for $\phi\rightarrow -\infty$, one finds that for $g(\phi)=e^{2\phi}Z(\phi)$,
\be
g(\phi) =1+\kappa_1\phi e^{2\phi}-(a-Z_m\ln Z_m)e^{2\phi}+O(e^{4\phi})\, .
\lb{105}
\ee
We see that only the physical fields that are in the Hartle-Hawking state contribute to the term $\phi e^{2\phi}$ in the metric function.
This is consistent with the fact that this term is due to the presence of the thermal radiation in the asymptotic region and that the thermal radiation is 
made of the physical particles only.

At the other end of the spacetime, when $\phi\rightarrow +\infty$, one finds that
\be
Z(\phi)=Z_\infty+\frac{Z_\infty}{Z_\infty+Z_m}e^{-2\phi}\, ,  
\lb{106}
\ee
where $Z_\infty$ is  the solution to equation $Z_\infty+Z_m\ln Z_\infty/Z_m=-a$. If $a\gg Z_m=\kappa_1/2$ one has that $Z_\infty=Z_m e^{-a/Z_m}$.
As a result, one finds the following expressions for the functions $g(\phi)$ and $h(\phi)$,
\be
g(\phi)=Z_\infty e^{2\phi}+\frac{Z_\infty}{Z_\infty+Z_m}\, , \  \ h(\phi)=-\lambda^{-1}\frac{Z_\infty}{Z_\infty+Z_m}\, .
\lb{107}
\ee
At this end, one has a singularity which is apparent from the resulting scalar curvature
\be
R=-{4\lambda^2}\frac{(Z_\infty+Z_m)^2}{Z_\infty}\, e^{2\phi}\, .
\lb{108}
\ee
Note, however, that unlike the singularity in the classical metric, where it is space-like,  this singularity is a time-like singularity.

As one varies $\phi$, the metric function $g(\phi)$ decreases from $g=1$ at $\phi=-\infty$ and then increases to infinity for $\phi=+\infty$. This indicates that $g(\phi)$ in-between  must have a minimum. This is indeed the case as can be seen by the analysis of the equation $g'(\phi)=0$. The minimum happens for $Z=Z_{min}$ and $\phi=\phi_{min}$ related by equation
$Z_{min}+Z_m=e^{-2\phi_{min}}$. The value of $Z_{min}$  and the minimal value of the metric function $g(\phi)$ are  found to be
\be
Z_{min}=(Z_m e) e^{-a/Z_m}\, , \ \   g(\phi_{min})=\frac{Z_{min}}{Z_{min}+Z_m}\, .
\lb{109}
\ee
One can check that $Z_{min}>Z_\infty$. When $a\gg Z_m$ one finds for the minimal value of the $(tt)$ component in the metric
\be
{\rm min} (-g_{tt})=e e^{-2a/\kappa_1}=e e^{-S_{BH}/\kappa_1}\, .
\lb{110}
\ee
It is exponentially small function of the entropy for the classical black hole. It is also non-perturbative function of $\kappa_1$. When $\kappa_1$ is taken to zero the minimal
value (\ref{110}) approaches zero.
The curvature is finite at the minimum,
\be
R=-4\lambda^2 e^{-1} e^{a/Z_m}\, .
\lb{111-1}
\ee
We note that when $\kappa_1\rightarrow 0$ the dilaton value at the minimum is moving to infinity, $\phi_{min}\rightarrow \infty$,  where  the curvature is divergent.  So that one can not interpret $\phi=\phi_{min}$ 
as the place where the classical horizon used to stay. It is rather the place where the singularity was located in the classical black hole solution.
On the other hand,  at the position of the classical horizon  $\phi=\phi_h=-1/2\ln a$ we find that $Z(\phi_h)=0.567 Z_m$ and hence the respective value of the metric function,
\be
g(\phi_h)=\frac{0.567 \kappa_1}{2a}=\frac{0.567\kappa_1}{S_{BH}}\, ,
\lb{111}
\ee
is bounded by the inverse  entropy of the classical black hole.  Clearly, when $\kappa_1\rightarrow 0$ one has that $g(\phi_h)=0$. 
This analysis shows that what used to be a horizon in the classical solution,  now becomes an extended region   between $\phi=\phi_h$  ($Z=0.567 Z_m$) to
$\phi=\phi_{min}$ ($Z=Z_{min}$) in the semiclassical solution.

In fact, we can also extend our computations for the general case of $\kappa_1\neq -\kappa_2$, where the total $\kappa=\kappa_1+\kappa_2$ may be positive or negative. Once again we have denoted physical fields with subscript 1 and ghosts with subscript 2. If we further assume $Z>0$ (which can be relaxed in a manner similar to subsection \ref{subsec:negZ}), then we have $A_1 = 0$ and $A_2 > 0$. Therefore $A = A_1+A_2 > 0$.

\medskip

\noindent\underline{\it The case of $\kappa>0$}: If we assume $\kappa>0$, then our master equation \eqref{103} takes the form 
\be\label{eq:hybrid1}
&&W(Z)=G(\phi)\, \qquad\text{with} \nonumber \\
&&  W(Z)=Z+A_2\ln\frac{Z}{A_2}\, \qquad\text{and} \qquad  G(\phi)=e^{-2\phi}+\kappa\phi-a\, ,
\ee
which is nothing but \eqref{eq:genAneg} studied before, with $A$ having the interpretation of $A_2$. Following our analysis there, we conclude that this hybrid case yields a spacetime with an asymptotically flat spacetime on one end and a naked singularity on the other, after passing through a throat of size $\sim 1/S_{BH}$.

\medskip

\noindent\underline{\it The case of $\kappa<0$}: On the other hand, if we have $\kappa<0$, then our master equation takes the form 
\be
&&W(Z)=G(\phi)\, , \nonumber \\
&&W(Z)=Z+A_2\ln\frac{Z}{A_2}\, ,  \  \  G(\phi)=e^{-2\phi}-k\phi-a\, ,
\lb{hybrid2}
\ee
which is what we encountered in \eqref{Z3}. As we know, the solution would therefore be asymptotically flat spacetimes at both ends with a throat which is exponentially small if expressed in terms of
 the classical black hole entropy (see \eqref{Z9} e.g.).

\medskip

Concluding this section let us note that the case of a hybrid quantum state is interesting in the following sense. This is the case where an outside
observer sees the thermal Hawking radiation of the physical particles at the classical Hawking temperature. On the other hand, the global spacetime does not have a horizon.
Thus, this is a quite unique example of co-existence of the known thermal properties of the classical black hole and a no-horizon semiclassical geometry. 
We remind that this situation happens when there is at least one non-physical quantum field (ghost) which is in the Boulware quantum state.
The sub-case of negative total $\kappa$ studied before is perhaps the most interesting since  the corresponding semiclassical spacetime is everywhere  non-singular.

\section{Concluding remarks}\label{sec:conclusions}
\setcounter{equation}0
\medskip

Finally, let us try to draw a few  obvious and less obvious conclusions.

\vspace{2mm}

\noindent\underline{\it Long throat picture}: \ \ What we find in the semiclassical RST model  is the following. The back-reacted geometry corresponding to  a generic quantum state different from 
the Hartle-Hawking state  is horizonless.  It appears that the Hartle-Hawking  is the only state for which the back-reacted geometry has a horizon and the entire space-time outside the horizon is a deformation of the classical black hole geometry. For all other quantum states, 
the classical horizon is replaced by a region which we call a throat, where the value of the $(tt)$ component in the metric
may be, depending on the value of the coupling parameter $\kappa$,  extremely small although non-zero. 
The ratio of $(-g_{(tt)})$  at the throat and at infinity defines a new time scale $t_P$ that tells us how slow is time running in the throat in comparison with the time at asymptotic infinity.
We have seen that $t_P$ is bounded by the inverse of the classical black hole entropy and in certain cases it can be exponentially small. This is consistent with the bound on the Poincar\'e recurrence
time discussed by Susskind \cite{Dyson:2002pf}, see also discussions in    \cite{BTZ,Kleban:2004rx,Barbon:2003aq,Germani:2015tda}.
Considering only the semiclassical solutions  which have a classical region,
the typical back-reacted geometry represents a spacetime which looks pretty close to the classical black hole up to a small region just outside the classical horizon.
Then, the horizon is replaced by a throat which may be quite long, again depending on the value of $\kappa$. On the other side of the throat one finds a spacetime with a singularity, either time-like or 
null-like.  The position of the singularity in the space is not dependent on the mass parameter $a$. 
Thus, in the space of the quantum states parameterized by real number $C$ (modulo the duality (\ref{22}) discussed in section \ref{sec:classandpolyakov}) there is only one point, $C=0$ (or $C=-4\lambda$)
for which the back-reacted geometry has a horizon. While for any other values of $C$ the geometry is horizonless with  a classical horizon being  replaced by  a throat. 

We also note that for some special cases,  such as studied respectively for the twisted solution in subsection \ref{subsec:7.3} and for the negative $Z$ case of subsection \ref{subsec:negZ}, the quantum spacetime doesn't have any asymptotically flat region. Rather it is bounded by null and naked singularities on either ends.

\vspace{2mm}

\noindent\underline{\it Black hole mimickers}: \ \ Each of the geometries for $C\neq 0, \, -4\lambda$ with a classical region gives us an example of a black hole mimicker.
Indeed, it behaves as the classical black hole geometry everywhere from the asymptotic infinity to the small region just outside the classical horizon that is now replaced by a throat.
The required travel time $t_H$ (see (\ref{70}))  for a light ray (sent from a point outside the throat)  to reach the center of the throat can be parametrically very large. For time observation much less than this characteristic time $t_H$,
no outside observer would be able to see any difference from the true black hole. It is important to note that what kind of   spacetime is on the other side of the throat plays no role 
in seeing this spacetime as a black hole mimicker: for times less than $t_H$ the part of the spacetime inside the throat is effectively cut off from the part outside the throat.
Thus, to be a mimicker, the spacetime does not have to be a wormhole as in \cite{Damour:2007ap} with two asymptotically flat regions.  It is in fact  sufficient to simply have a throat with large characteristic time $t_H$. This opens up a bigger class of geometries that may represent the black hole mimickers. 

\vspace{2mm}

\noindent\underline{\it Consequences for information puzzle}:  \ \ As we see in the present analysis of a generic quantum state (apart from the Hartle-Hawking one), it
is represented by a horizonless spacetime. It is sufficient to just have one quantum field in the Boulware state, for the entire classical horizon to disappear.
The Boulware state is the only physically meaningful  quantum state for the non-physical fields such as ghosts. Generically ghosts are ubiquitous, as they appear either when the gauge fields are quantized or 
when the gravity itself is quantized. In quantum field theory described by a unitary $S$-matrix, the ghosts are not present in the asymptotic states although they may appear in the 
intermediate interactions  deep in the bulk of spacetime.  For example, in the presence of gravity, the ghosts should not have a non-vanishing stress tensor to be detected at the asymptotic infinity
although it may be non-zero somewhere in the bulk of spacetime.
This uniquely singles out the Boulware state for the ghosts. Hence, in the presence of ghosts even if all physical fields are in the Hartle-Hawking state the classical horizon is removed and is
replaced by a throat.  The information paradox is usually formulated in terms of the classical black hole spacetime with a horizon that is formed in the process of the gravitational collapse.
That is a dynamical process which we did not analyze in the present paper. However, it seems  quite natural to expect  that the static geometries are good approximation for a dynamical, slowly evolving situation
so that the time evolution of the metric can be thought of as a slow passage from one static geometry to the other. The absence of a horizon in the semiclassical geometry means that the paradox  vanishes in any practical sense. The problems related to the loss of information inside the horizon in the classical picture now becomes a better defined problem of the information passing through the long throat and the problem of interaction with the singularity. 
The long delay in the possible retrieval of the information sent into ``black hole''  due to its long passing in the throat mimics the information loss in the sense that
the information appears to be lost for any observation time much less than  $t_H$ while fundamentally no actual loss happens. 
On the other hand, the interaction with the singularity is a new problem that arises. However, it might be treated in a rather conventional way.
For instance, the presence of a time-like singularity    plays a role similar to the boundary for a quantum field and it simply requires a formulation of certain ``boundary conditions'' at the singularity (see e.g. discussions in \cite{Horowitz:1995gi}).  We note that the singularity related problems are absent in a hybrid scenario when $\kappa$ is negative in which case the spacetime is everywhere regular.

Finally, given our long-throat geometries arise quite universally, such spacetimes seem to be very naturally present as saddle points of some exact, quantum corrected actions (as we also saw in four-dimensions in \cite{Berthiere:2017tms}). Existence of such horizonless saddles should play a pivotal role in the information loss problem (via the approach of quantum gravity path integral). It will also be interesting to see whether our findings here have any connections with the recent developments in the information puzzle reviewed in \cite{Almheiri:2020cfm}.  In future we want to study these and the related issues more carefully.

\vspace{0.2 cm}
\centerline{\bf Acknowledgements}

\noindent SS would  like to thank T.~Jacobson, E.~Gourgoulhon, T.~Wiseman and O.~Zaslavsky for useful communications and discussions. DS would like to thank the CHEP group at Indian Institute of Science (IISc.) for their kind hospitality and support during the final part of this project.

\end{document}